\documentclass[showpacs,showkeys,prc,aps,preprintnumbers]{revtex4}
\usepackage{graphicx}

\newcommand{ \be }{\begin{equation}}
\newcommand{ \ee }{\end{equation}}
\newcommand{ \bea }{\begin{eqnarray}}
\newcommand{ \eea }{\end{eqnarray}}
\newcommand{ \la }{\langle}
\newcommand{ \ra }{\rangle}
\newcommand{ \lla }{\left \langle}
\newcommand{ \rra }{\right \rangle}

\newcommand{ \vphi }{\varphi}
\newcommand{ \vphii }{\varphi _i}
\newcommand{ \vphij }{\varphi _j}
\newcommand{ \vphik }{\varphi _k}
\newcommand{ \si }{\sigma _i}
\newcommand{ \sj }{\sigma _j}
\newcommand{ \sk }{\sigma _k}
\newcommand{ \sij }{\sigma _{ij}}

\usepackage{color}

\begin{document}

\title{ Three-particle cumulant Study of Conical Emission}
\begin{flushleft}  \small \sl version 3.7,  \today \\  \end{flushleft} 
\author{Claude A. Pruneau}
\affiliation{Physics and Astronomy Department, Wayne State University, 
Detroit, MI 48201 USA}

\begin{abstract}
We discuss the sensitivity of the three-particle azimuthal cumulant method for a search and study 
of conical emission in central relativistic $A+A $ collisions. Our study is based on a multicomponent Monte Carlo model
which include flow background, Gaussian mono-jets, jet-flow, and Gaussian conical signals. We find the observation
of conical emission is hindered by the presence of flow harmonics of fourth order ( $v_4 $) but remains feasible
even in the presence of a substantial background. We consider the use of probability cumulants for the suppression
of 2$^{nd}$ order flow harmonics. We find that while probability cumulant significantly reduce  $v_2^2$ contributions, they also
complicate the cumulant of jets, and conical emission. The use of probability cumulants is therefore not particularly advantageous in searches for 
conical emission. We find the sensitivity of the (density) cumulant method depends inextricably on strengths of 
 $v_2 $,  $v_4 $, background and non-Poisson character of particle production. It thus cannot be expressed in a simple 
form, and without specific assumptions about the values of these parameters. 
\end{abstract}

\maketitle

\section{Introduction}
\label{Sect:Introduction}

Observations of away-side dip structures in two-particle correlations measured in $Au+Au$collisions at 
 $\sqrt{s_{NN}} = 200 $ GeV have stimulated renewed interest in the notion of conical emission. The 
passage of a parton through dense matter at speeds greater than the speed of sound is predicted to lead 
to the production of a Mach shock wake resulting in conical emission pattern that may explain the 
observed two-particle correlations \cite{Stoecker76,Stoecker05,JRuppert05,RupperMuller05,SolanaShuryak05,RenkRuppert06,Heinz06,Stocker07,RenkRupper07,MullerQM08,BBetz08}. 
Identification of Mach cone is of great interest because it could provide an 
experimental determination of the speed of sound in the dense medium produced in high energy  $A+A $ collisions 
\cite{SolanaShuryak05,Mukherjee07}. A sonic boom is also expected for a heavy quark propagation 
based on Ads/CFT calculation \cite{Gubser08}. Cerenkov radiation produced by a superluminal parton 
traversing a dense medium is expected to generate a similar signature 
\cite{Dremin79,Dremin81,Dremin06,KochMajumber05,MajumderMuller06}. However, other 
production mechanisms have been proposed to explain the two-particle correlation data. These include 
large angle gluon radiation \cite{Vitev05,PolosaSalgado05}, path length effects \cite{Hwa05}, 
collective flow, and  jet deflection \cite{Armesto04,Salgado05a,RenkRuppert07_PLB646,RenkRuppertPRC2007,Voloshin05,Hwa05_PRC74}.

Three-particle correlation measurements were proposed to gain further insight in the particle production 
mechanism leading to the away-side deep structure seen in two-particle correlations. Various methods 
have been suggested and are currently pursued to carry such analyses.
A number of these analyses are carried out using a flow $+$ signal decomposition, i.e. an adhoc flow background
is assumed and subtracted based on the ZYAM 
approximation \cite{Holzmann05,Ajitanand06,Ulery06_774,Ulery06a,Ulery06b,Ulery06c,Ajitanand06_774,Ulery07,CeresQM06,Kniege07,Ajitanand07}.
Note however that, as Borghini pointed out, momentum conservation can have a significant impact on two- and three-particle correlations \cite{Borghini06}. 
The magnitude of momentum conservation effects is however difficult to estimate as discussed recently by Chajecki {\it et al.} \cite{Lisa08}.
A cumulant based method was proposed by the author \cite{Pruneau06} and is also used by the STAR collaboration \cite{Pruneau06a} to search 
for conical emission. This method presents the advantage that the extraction of a three-particle signal does {\it not} require any model assumptions.
Obviously, one can also carry a model based decomposition of the extracted three-particle signal, i.e. the cumulant, to estimate its various components, 
and seek evidence for conical emission.
 
In this work, we further study the properties of the three-particle cumulant observable first proposed in 
Ref. \cite{Pruneau06} for a search for conical emission, and studies of particle production dynamics. 
Three particles correlations, designed to search for conical emission, use a high  $p_t $ particle as a jet tag. 
The azimuthal angle of this high  $p_t $ particle should approximately correspond to the direction of the 
parton initiating the tagged jet. It is further assumed the emission of this parton is surface
biased, and directed outward, approximately normal to the surface of the medium. The direction of the high-$p_t$
particle thus provides a reference to study the propagation of the away-side parton initiating the second
jet. Estimates of the sound velocity in the quark gluon plasma suggest the wake produced by a high energy 
parton should produce particle emission at a Mach angle of the order of one radian relative to the parton 
direction \cite{SolanaShuryak05}.  This corresponds to enhanced particle emission at  angles of order 2 
radian with respect to the high-$p_t$ jet tag.  An unsuppressed away-side jet should on the other hand 
lead to particle emission at an angle of approximately  $\pi $ relative to the jet tag.  Deflection or radial 
flow effects should produce a broadening of the away-side jet relative to the high-$p_t$ tag.


We define the notation and variables used in this paper in Section \ref{Sect:definitions}. We next discuss, in Section 
\ref{Sect:SignalModeling}, simple models of particle production including di-jets, flow, conical emission, as
well as the effects of differential jet quenching, hereafter 
called jet-flow.  We show in Sec. \ref{Sect:flow}
that in the case of Poisson statistical particle production, the three particle cumulant 
(hereafter noted {\it 3-cumulant})  associated 
with azimuthal anisotropy  reduces to a simple expression involving only non-diagonal Fourier 
terms. In view of the large  values of elliptic flow  $v_2  $
observed in  $A+A $ collisions at RHIC, one expects the leading non-diagonal term should be of order 
 $v_2 v_2 v_4$. However, one finds experimentally that fluctuations in the number of produced particles are 
non-Poissonian. We thus consider in Sec. \ref{Sect:ProbabilityCumulants} the impact of such non-Poissonian 
fluctuations, and find they imply the presence of sizable  $v_2 ^2  $
 terms in the three-cumulant that may complicate the observation of conical emission signals. We show 
the  $v_2 ^2  $
terms may however be suppressed if one adopts a modified version of the cumulant based on 
probability densities rather than number densities, and suggest this modified cumulant as an alternative means of studying three-particle correlations.   

The cumulant method is a relatively complicated analysis technique, which may in principle be 
sensitive to various instrumental effects. We discuss a specific implementation of the method based on 
normalization to single particle distributions in Sec. \ref{Sect:EfficiencyCorrection}. We show, based on the conical emission model introduced 
in Sec. \ref{Sect:Sensitivity}  and an adhoc parameterization of the detection efficiency, that 
normalization by single particle distributions leads to a robust analysis technique.
 
A key issue in the search for conical emission is the sensitivity of the method used to extract the three-
particle correlation information. We discuss, in Section \ref{Sect:Sensitivity}, the sensitivity of the 
cumulant method based on a simple jet and conical emission Monte-Carlo model. 

Our conclusions are summarized in Section \ref{Sect:Conclusion}.

\section{Definitions and Notation}
\label{Sect:definitions}

The conical emission search method introduced in Ref. \cite{Pruneau06} is based on the observation 
of three-particle densities as a function of the relative azimuthal angles between the measured particles. 
The presence of three-particle correlations, and possible signal for conical emission is extracted using  
the {\it 3-cumulants}. A high  $p_t $ particle is used as a jet tag, and proxy for 
determination of the direction of the jet. Two lower  $p_t $ particles are used to probe the structure of the 
near-side jet (i.e. the jet singled out by the high-$p_t$ tag particle) and search for conical emission on the 
away-side.  The jet tag is herein referred as particle 1 while the two associates are labeled as particles 2 
and 3. The three particles are detected in the collision transverse plane at angles  $\vphi_1 $, 
 $\vphi_2 $,  and  $\vphi_3 $ relative to some arbitrary reference frame.  The single, two-, and three-
particle densities are noted as follows:
\begin{eqnarray}
\rho_1 (\vphii ) &=& dN_1 /d\vphii  \\
\rho_2 (\vphii,\vphij ) &=& dN_2 /d\vphii d\vphij  \nonumber \\  
\rho_3 (\vphii,\vphij,\vphik ) &=& dN_3 /d\vphii d\vphij d\vphik \nonumber
\label{densities}
\end{eqnarray}
where the indices  $i $,  $j $, and  $k $ take values 1, 2, and 3. 

The three-particle or triplet density, $\rho_3$, corresponds to the average number of particle triplets observed per 
collision. The particles of a given triplet are however not necessarily correlated. Indeed, the three 
particles may originate from one, two, or three (distinct or not) production processes (e.g. radial flow, 
elliptic flow, resonance decay, jets, Mach cone, etc). Cumulants are designed to extract the three-particle correlation 
component from the three-particle density. They were first discussed in the context 
of particle physics (see for instance \cite{Berger,Carruthers}) and are now used in a 
variety of analyses.  It is the purpose of this work to study 
specific aspects of the cumulant method introduced in \cite{Pruneau06} for studies of the shape and strength of the signal expected from 
different types of processes and to characterize the robustness and sensitivity of the method.

The 2- and 3-cumulants  $C_2$ and  $C_3$ are defined as follows:
\begin{eqnarray}
 C_2(\vphii,\vphij )        &\equiv&\rho_2(\vphii,\vphij)-\rho_1 (\vphii )\rho_1 (\vphij ) \nonumber \\ 
 C_3(\vphii,\vphij,\vphik ) &\equiv&\rho_3(\vphii,\vphij,\vphik ) - \rho_2(\vphii,\vphij )\rho_1 (\vphik )\\
                            &  -   &\rho_2(\vphii,\vphik)\rho_1 (\vphij ) - \rho_2 (\vphij,\vphik )\rho_1 (\vphii ) \nonumber \\ 
                            &  +   &2\rho_1 (\vphii )\rho_1 (\vphij )\rho_1 (\vphik )  \nonumber
\label{cumulants}
\end{eqnarray}
As described in Ref. \cite{Pruneau06} the two-, three-particle densities and cumulants are straightforwardly 
corrected for detector inefficiencies provided the three- and two-particle detection efficiencies may be 
factorized as products of respectively two and three single particle efficiencies. The robustness of this correction procedure is 
discussed in Sec. \ref{Sect:EfficiencyCorrection} on the basis of Monte Carlo simulations.

Correlation functions in terms of relative angles,  $C_3(\Delta \vphi_{ij},\Delta \vphi_{ik} ) $, are formally obtained by 
integration of the cumulants  $C_3(\vphii,\vphij,\vphik) $ with constraints  $\Delta\vphi_{ij}=\vphii-\vphij $.
\bea
C_3(\Delta\vphi_{ij},\Delta\vphi_{ik}) &=& \int{C_3(\vphii,\vphij,\vphik)} \\ 
 &\times& \delta(\Delta\vphi_{ij}-\vphii+\vphij)\nonumber \\
 &\times& \delta(\Delta\vphi_{ik}-\vphii+\vphik)d\vphii d\vphij d\vphik \nonumber
 \label{cumulantConstraint}
\eea
In practice, this is accomplished by binning  $C_3 (\vphii,\vphij,\vphik )  $ and  $C_3 (\Delta \vphi_{ij},\Delta \vphi_{ik} ) $ into arrays (e.g. 72x72x72 and 72x72 respectively) and summing 
the elements of  $C_3 (\vphii,\vphij,\vphik )  $  to obtain  $C_3 (\Delta \vphi_{ij},\Delta \vphi_{ik} )  $
as follows. 
\be
C_3(p,q)=\sum\limits_{i,j,k = 1}^{72}C_3 (i,j,k) \delta (p - i + k)\delta (q - i + k)
\label{cumulantDiscrete}
\ee
for  $p $=1, ...,72;  $q $=1, ...,72. Given the finite statistics, and the large memory requirement implied by three-dimensional arrays used 
in this type of analysis, care must be taken when binning the densities. We found the use of 72 bins, for 
analysis of data samples collected e.g. by the STAR experiment \cite{StarNim,StarC308} enables sufficient angular resolution and statistical accuracy.

\section{Signal Modeling}
\label{Sect:SignalModeling}

We use specific models for jet, di-jet production, collective flow, and conical emission signals in 
order to assess the sensitivity of the cumulant method for a search of conical emission.  The models are 
described in the following  sub-sections while our study of the sensitivity of the cumulant method is presented 
in Section \ref{Sect:Sensitivity}. 

\subsection{Di-Jet Production}
\label{Sect:DiJetProduction}

\begin{figure}[!htP]
\mbox{
  \begin{minipage}{0.49\linewidth} 
  \begin{center}
  \includegraphics[width=0.99\linewidth]{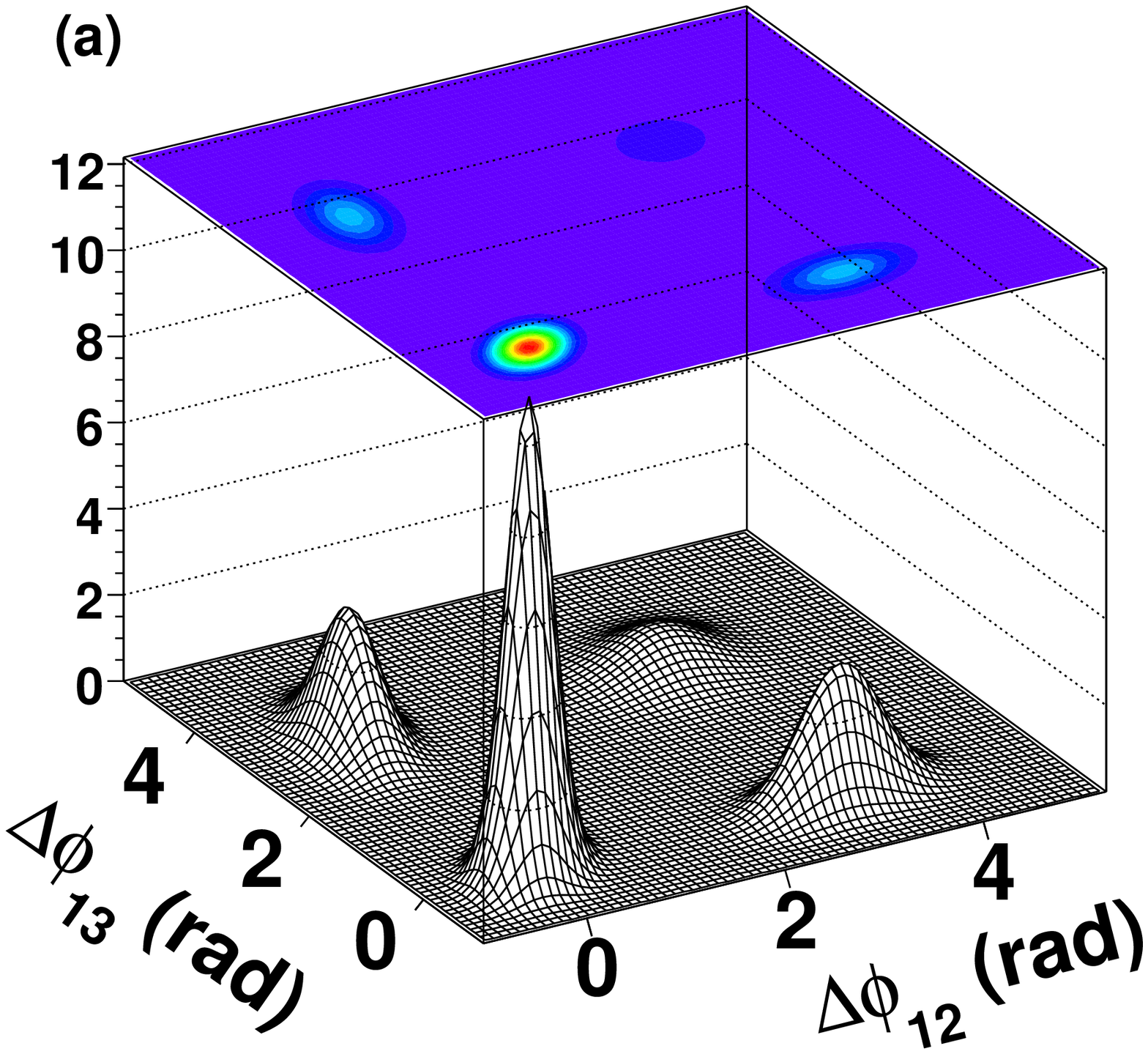}
  \end{center}
  \end{minipage}  
  \begin{minipage}{0.49\linewidth}
  \begin{center}
  \includegraphics[width=0.99\linewidth]{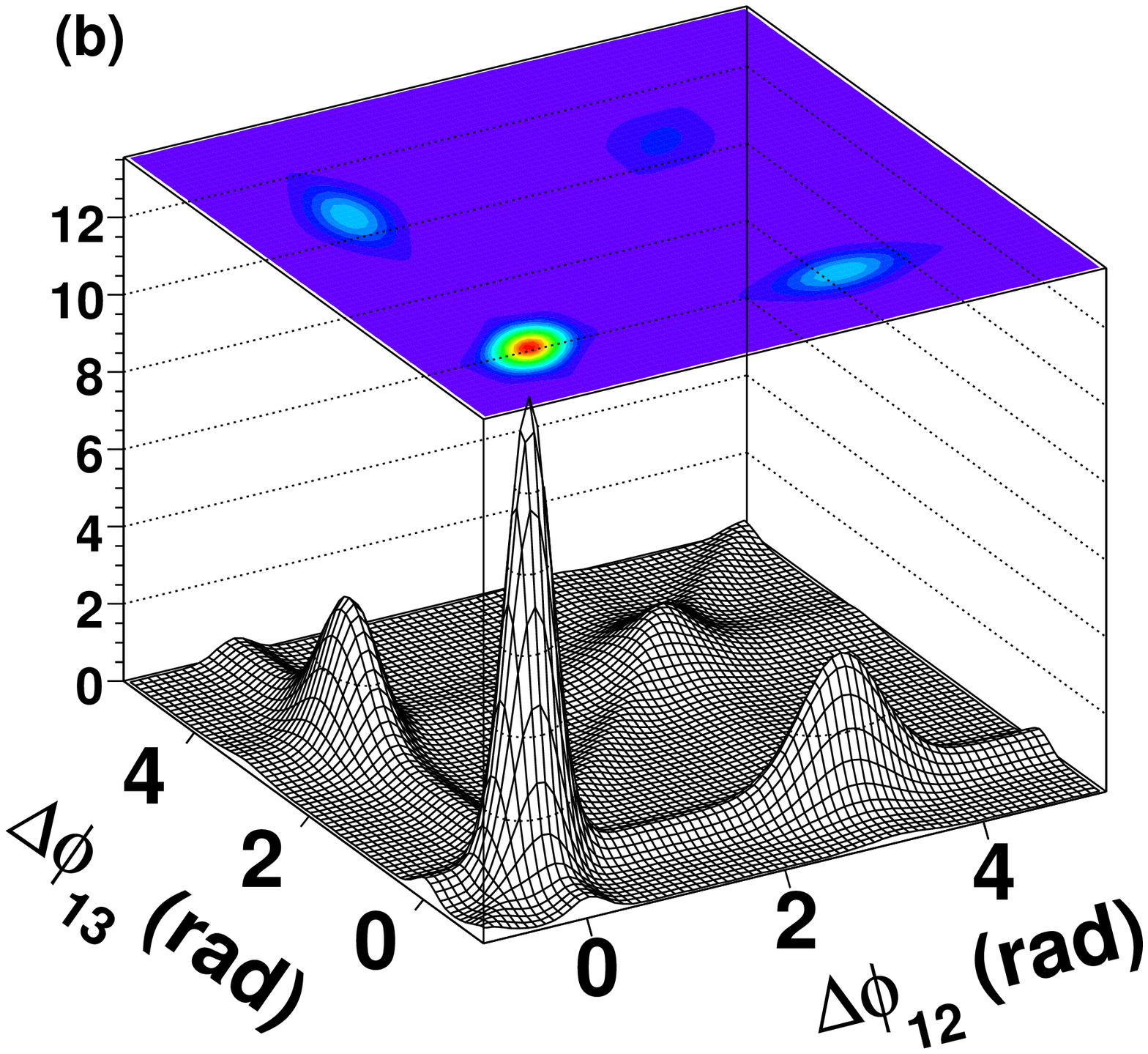}
  \end{center}
  \end{minipage}
}
\caption{3-Cumulants calculated based on the Gaussian jet model discussed in the text. (a) Poisson case, 
(b) Non-Poisson case with  $\la J(J-1) \ra - \la J \ra^2 = 0.05\la J \ra $}
\label{fig:gauss}
\end{figure}

Jet production is characterized by the emission of particles in a cone (in momentum space) centered on 
the direction of the parton that produces the jet. We consider di-jet production restricted to central 
rapidities (near  $90^o $ relative to the beam direction), and assume the number of di-jet per event in the 
acceptance of the detector, J, varies event-by-event. The intra-jet particle multiplicity depends on 
the jet energy, but is order three (3) at RHIC  energy \cite{CDF_JET}. We consider back-to-back jets in 
azimuth, but possibly different longitudinal momenta. We denote the number of jet particles measured 
within each kinematics cut "$i$" as  $A_i $. and  $A_i'$ respectively for the near and away side jets. The emission 
of particles relative to the jet axis is described by a probability distribution  $P_{assoc} (\theta ) $,
 where  $\theta  $ is the angle between the particle momentum and the momentum vector of the parton originating the jet. 
Projection of this distribution in the transverse plane leads to a probability distribution,  $P_J (\vphi ) $,
 function of the azimuthal angle,  $\vphi  $, between the parton and the particle direction.  Following  \cite{Pruneau06}, we further simplify the jet 
model and use a Gaussian azimuthal profile,  $G_2 (\vphii  - \phi_\alpha  ;\si ) $
with  $G_2(x;\sigma ) \equiv \left( {\sqrt {2\pi } \sigma } \right)^{ - 1} \exp \left( { - {{x^2 } \mathord{\left/
 {\vphantom {{x^2 } {2\sigma ^2 }}} \right. \kern-\nulldelimiterspace} {2\sigma ^2 }}} \right) $
 where  $\vphii  $,  $\phi_\alpha   $, and  $\si  $ are the emission angle (in the lab frame) of measured particles, the emission angle of the 
parton, and the width of the jet, respectively. For this illustrative model, we assume the emission angle and the jet 
multiplicity are not correlated to the collision reaction plane. We further assume one can decouple the 
number of particles, in the measured kinematic range of interest, from the jet profile. The number of 
jets, their multiplicity, and directions are not known, and must therefore be averaged out. The 3-particle 
jet cumulant is given by the following expression.
\bea
 C_3^{Jet} (\Delta \vphi_{12},\Delta \vphi_{13} ) &=& (2\pi)^{-1} \lla J \rra \left\{ 
\begin{array}{l}
 \lla {A_1 A_2 A_3 } \rra G_3 (\Delta \vphi_{12},\Delta \vphi_{13} ;\sigma_1,\sigma_2,\sigma_3 ) \\ 
  + \lla {A_1 A_2 A'_3 } \rra G_3 (\Delta \vphi_{12},\Delta \vphi_{13}  - \pi ;\sigma_1,\sigma_2,\sigma '_3 ) \\ 
  + \lla {A_1 A'_2 A_3 } \rra G_3 (\Delta \vphi_{12}  - \pi,\Delta \vphi_{13} ;\sigma_1,\sigma '_2,\sigma_3 ) \\ 
  + \lla {A_1 A'_2 A'_3 } \rra G_3 (\Delta \vphi_{12}  - \pi,\Delta \vphi_{13}  - \pi ;\sigma_1,\sigma '_2,\sigma '_3 ) \\ 
 \end{array} \right\} \\ 
 & +& (2\pi)^{-2} \left( {\lla {J(J - 1)} \rra  - \lla J \rra ^2 } \right)\left\{ 
\begin{array}{l}
 \lla {A_1 A_2 } \rra \lla {A_3  + A'_3 } \rra G_2 (\Delta \vphi_{12} ;\sigma_{12} ) \\ 
  + \lla {A_1 A_3 } \rra \lla {A_2  + A'_2 } \rra G_2 (\Delta \vphi_{13} ;\sigma_{13} ) \\ 
  + \lla {A_2 A_3 } \rra \lla {A_1 } \rra G_2 (\Delta \vphi_{13}  - \Delta \vphi_{12} ;\sigma_{23} ) \\ 
  + \lla {A_1 A'_2 } \rra \lla {A_3  + A'_3 } \rra G_2 (\Delta \vphi_{12}-\pi;\sigma'_{12} )\\ 
  + \lla {A_1 A'_3 } \rra \lla {A_2  + A'_2 } \rra G_2 (\Delta \vphi_{13}-\pi;\sigma'_{13} )\\ 
  + \lla {A'_2 A'_3 } \rra \lla {A_1 } \rra G_2 (\Delta \vphi_{13}-\Delta \vphi_{12} ;\sigma '_{23} )\\ 
 \end{array} \right\} \nonumber\\ 
 & +& (2\pi)^{-3}\left( {\lla {J(J - 1)(J - 2)} \rra  - 3\lla {J(J - 1)} \rra \lla J \rra  + 2\lla J \rra ^3 } \right)\lla {A_1 } \rra \lla {A_2 + A'_2 } \rra \lla {A_3  + A'_3 } \rra  \nonumber
\label{cumulantJet}
\eea
with
\be
G_3 (x_1,x_2 ;\sigma_1,\sigma_2,\sigma_3 ) = \left(2\pi\right)^{-1}\sigma_{1,2,3}^{-2} \\
\times\exp\left({-{{\left({\sigma_3^2 x_1^2+\sigma_2^2 x_2^2+\sigma_1^2 (x_1-x_2 )^2}\right)}\mathord{\left/{\vphantom{{\left( {\sigma_{_3 }^2 x_1 ^2  + \sigma_{_2 }^2 x_2 ^2  + \sigma_{_1 }^2 (x_1  - 
x_2 )^2 } \right)} {2\sigma_{1,2,3} ^2 }}} \right.} {2\sigma_{1,2,3}^2}}} \right)
\label{g3}
\ee
where  $x_{ij}=x_i-x_j $,  $\sij^2=\si^2+\sj^2 $ and  $\sigma_{i,j,k}^4=\si^2 \sj^2  + \si^2 \sk^2  + \sj^2 \sigma _k^2  $.
Note $G_3$ is defined such its integral over $x_1$ and $x_2$ is unity.
The coefficient  $\lla J \rra  $
determines the average number of jets per event. Coefficients  $A_1 $,  $A_2 $,  $A_3 $ describe the number of 
particles associated with the near side jet, while  $A'_1 $,  $A'_2 $,  $A'_3 $ correspond to the number of particles 
associated with the away side jet observed within the kinematic cuts used for the study of the 
correlation functions. We note that if the number of jets, J, in each event is determined by a Poisson process, then one has 
 $\lla {J(J - 1)} \rra  = \lla J \rra ^2  $ and $\lla {J(J - 1)(J - 2)} \rra  - 3\lla {J(J - 1)} \rra  - 2\lla J \rra ^3  = 0 $.
The constant term, and the terms containing a 2-particle dependence in  $G_2  $ then vanish in the above expression.
The jet 3-cumulant is plotted in Figure 1, in arbitrary amplitude units, 
for $\sigma_1  = 0.15 $,  $\sigma_2  = \sigma_3  = 0.2 $, $\sigma'_2  = \sigma'_3  = 0.4 $, $\lla J \rra  = 1 $,  $A_1  = 1 $,
  $A_2  = A_3  = 2 $, and $A'_2  = A'_3  = 1 $. We further assume for simplicity that 
  $\lla A_i A_j\rra = \lla A_i\rra\lla A_j\rra$
  and  $\lla A_1A_2A_3\rra = \lla A_1\rra\lla A_2\rra\lla A_3\rra$ although it is unlikely realized in practice.
   Panel (a) presents the cumulant for the Poissonian case, whereas panel (b) displays a 
 non-Poissonian case for which  ${{\lla {J(J - 1)} \rra  - \lla J \rra ^2 } 
\mathord{\left/ {\vphantom { \left( {\lla {J(J - 1)} \rra  - \lla J \rra ^2 } \right) {\lla J \rra ^2 }}} \right.
 \kern-\nulldelimiterspace} {\lla J \rra ^2 }} = 0.05 $. The peak centered at  $\Delta\vphi_{12}=\Delta\vphi_{13}=0 $
 corresponds to two particles associated with the high  $p_t $ particle, and is referred to as the near side 
jet peak. Secondary peaks at  $(0,\pi ) $,  $(\pi,0) $, and  $(\pi,\pi ) $ correspond to one or both the associates being detected on the away side. 
The bands seen in Figure \ref{fig:gauss} (b) stem from the non-zero two-particle contributions to the 3-cumulant for non-Poissonian events. 
Various effects may alter the strength and shape of the jet correlations. Interactions of the parton with 
the medium may produce jet broadening, and deflection. Given jet emission is expected to be surface 
biased in $A+A$ collisions (in view of recent measurements of small  $R_{AA} $, and the disappearance of the away side jet 
\cite{WhitePapers}), one expects the near side peak to have similar width in  $A + A$ 
collisions as in  $p + p$ collisions, while the away-side peak should be broader. One might also observe 
additional broadening along  $\Delta \vphi_{12}  $, and  $\Delta \vphi_{13}  $ due to parton scattering, and radial flow. 

\subsection{Collective Flow}
\label{Sect:flow}

Flow, or collective motion, is an important feature of heavy ion collisions at relativistic energies. It 
manifests itself by a modification of transverse momentum ($p_t$) spectra relative to those observed in $p+p$ collisions, and by azimuthal 
anisotropy of produced particles. In this section, we focus on 
azimuthal anisotropy arising in non-central heavy ion collisions. We decompose the azimuthal 
anisotropy in terms of harmonics relative to an assumed reaction plane. The conditional probability, $P_F (\vphii |\psi )$, to 
observe a particle at a given azimuthal angle,  $\vphii $, with the reaction plane at angle,  $\psi $, is written as a Fourier series. 
\be
P_F (\vphii |\psi )=F_2(\vphii-\psi;v_m(i)).
\label{flowFourier}
\ee
where
\be
F_2(\Delta\vphi;v_m ) = 1 + 2\sum\limits_m^{}{v_m}\cos \left( {m\left( {\Delta\vphi } \right)} \right)
\label{Fourier}
\ee
The Fourier coefficients  $v_m(i)$ measure the m-th order anisotropy for particles emitted in a selected kinematic range $i$. 
Measurements have shown the second order (elliptical) anisotropy can be rather large in  $Au + Au $ 
collisions at RHIC while first and fourth order harmonics are typically much smaller. STAR 
measurements show the fourth harmonic scales roughly as the square of the second order harmonic 
($v_4  \approx 1.1v_2^2 $) \cite{WhitePapers,StarV4}. The third and fifth harmonics are by symmetry null at  $\eta=0 $ for a colliding system such as 
 $Au + Au $, and expected to be rather small at other rapidities. Higher harmonics are most likely 
negligible. We neglect possible event-to-event fluctuations of these coefficients given it emerges from
recent works disentangling flow fluctuations and non-flow correlations is difficult. We describe 
the probability of finding  $N_i $ particles in the kinematical range $i$ according to probability  $P(N_i) $. The 
exact form of this probability is not required. Only its first, second, and third moments are needed. The 
joint probability of measuring  $N_i $ particles at an angle  $\vphii  $
while the reaction plane angle is at  $\psi  $
 is given by  $ P_F (\vphii,N_i,\psi ) = P_F (\vphii |\psi )P_F (N_i )P_{RP} (\psi ) $
 where  $P_{RP} (\psi ) = (2\pi )^{ - 1}  $
 is the probability of finding the reaction plane at a given angle  $\psi $. Integration yields the single particle 
density  $\rho_1 (\vphii )= \left( {2\pi } \right)^{ - 1} \lla {N_i } \rra  $.
The flow 2- and 3-cumulants are given respectively by (see Ref.\cite{Pruneau06} for a derivation of these 
expressions):
\bea
C_2^F(\Delta\vphi_{ij})&=&\left(2\pi\right)^{-2} \left( \lla N_i N_j\rra F_2(\Delta\vphi_{ij};v_m (i)v_m (j)) -\lla N_i \rra\lla N_j\rra \right) \nonumber\\
                       &=&\left(2\pi\right)^{-2}\lla N_i N_j\rra \\
                       & & \times \left({1-d_{ij}+ 2\sum\limits_m^{} {v_m (i)v_m (j)\cos\left( {m\Delta \vphi_{ij} } \right)} } \right)\nonumber
\label{c2Flow}
\eea
and
\begin{eqnarray}
C_3^F(\vphii,\vphij,\vphik ) &=& (2\pi)^{-3}\lla N_1 N_2 N_3\rra \left[ \Phi_3(\vphii,\vphij,\vphik) \right. \\ 
  &+& \left(1-f_{ijk}\right)\Phi_2(\Delta\vphi_{ij}) \nonumber \\ 
  &+& \left(1-f_{ikj}\right)\Phi_2(\Delta\vphi_{ik}) \nonumber \\ 
  &+& \left(1-f_{jki}\right)\Phi_2(\Delta\vphi_{jk}) \nonumber \\ 
  &+& \left. 1 - f_{ijk}  - f_{ikj}  - f_{jki}  + 2g_{ijk} \right] \nonumber
  \label{c3Flow}
\end{eqnarray}
where 
\be
\Phi_2 (x;v_{1,m},v_{2,m} )=2\sum\limits_{m = 1}^\infty  {v_{1,m} v_{2,m} \cos (mx)} 
\label{phi2}
\ee
and
\bea
\Phi_3 (\vphii,\vphij,\vphik) = 2\sum\limits_{p,m,n}^{} {v_{1,p} v_{2,m} v_{3,n} \left[ \begin{array}{l}
 \delta_{p,m + n} \cos \left( {p\vphii  - m\vphij  - n\vphik } \right) \\ 
  + \delta_{m,p + n} \cos \left( { - p\vphii  + m\vphij  - n\vphik } \right) \\ 
  + \delta_{n,m + p} \cos \left( { - p\vphii  - m\vphij  + n\vphik } \right) \\ 
 \end{array} \right]}
 \label{phi3}
\eea	
 $v_{i,m} $ correspond to the anisotropy coefficients of order $m$ for particle $i$. The coefficients  $d_{ij} $,  $f_{ijk} $, and  $g_{ijk} $ 
are defined as follows
\bea
 d_{ij}  &=& \frac{{\lla {N_i }     \rra \lla {N_j } \rra }}{{\lla {N_i N_j } \rra }} \nonumber \\ 
 f_{ijk} &=& \frac{{\lla {N_i N_j } \rra \lla {N_k } \rra }}{{\lla {N_i N_j N_k } \rra }} \\ 
 g_{ijk} &=& \frac{{\lla {N_i }    \rra \lla {N_j } \rra \lla {N_k } \rra }}{{\lla {N_i N_j N_k } \rra }} \nonumber 
 \label{dfg}
\eea
\begin{figure}[!htP]
\mbox{
  \begin{minipage}{0.5\linewidth} 
  \begin{center}
  \includegraphics[width=0.99\linewidth]{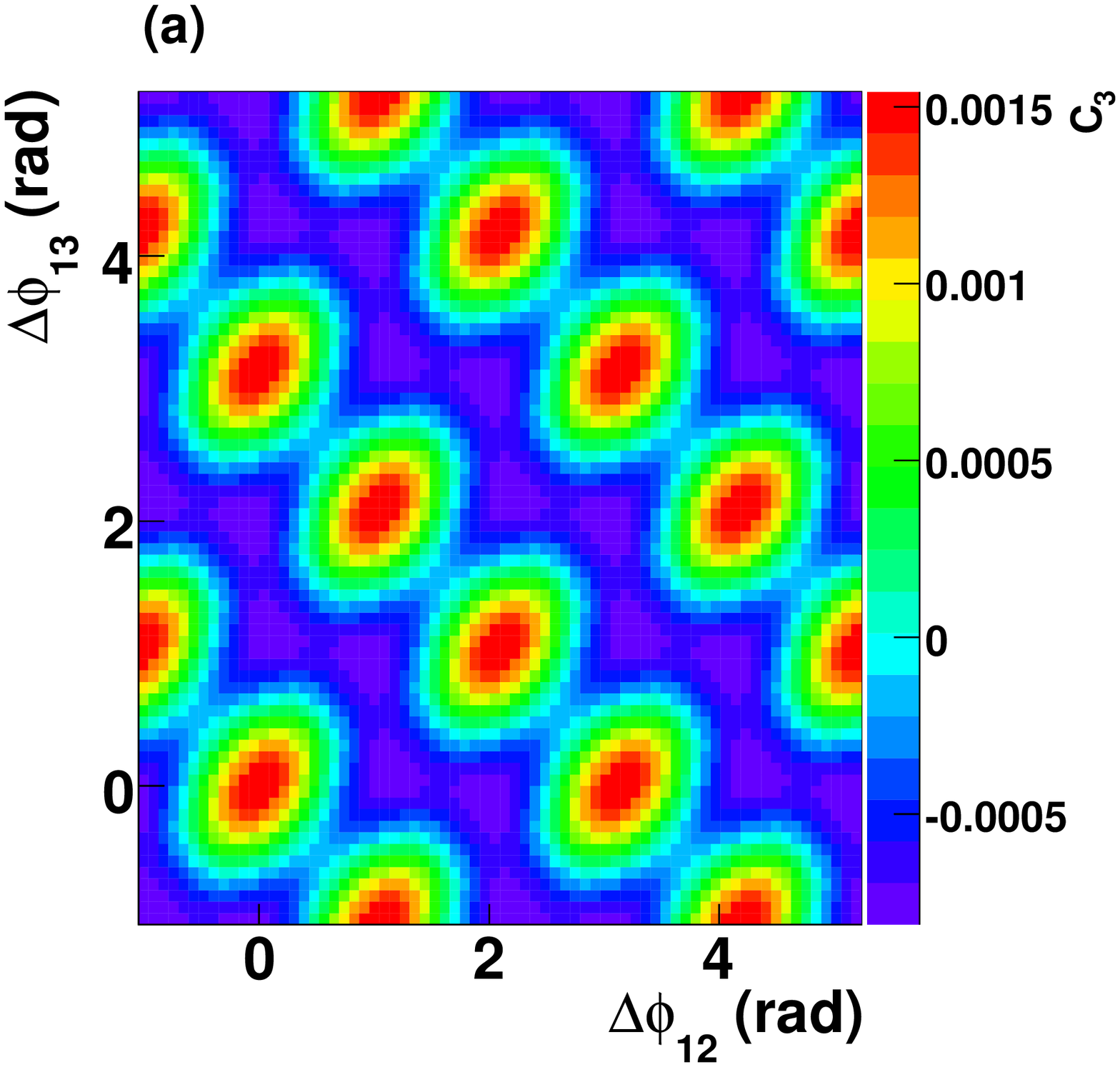}
  \end{center}
  \end{minipage}
  \begin{minipage}{0.5\linewidth}
  \begin{center}
  \includegraphics[width=0.99\linewidth]{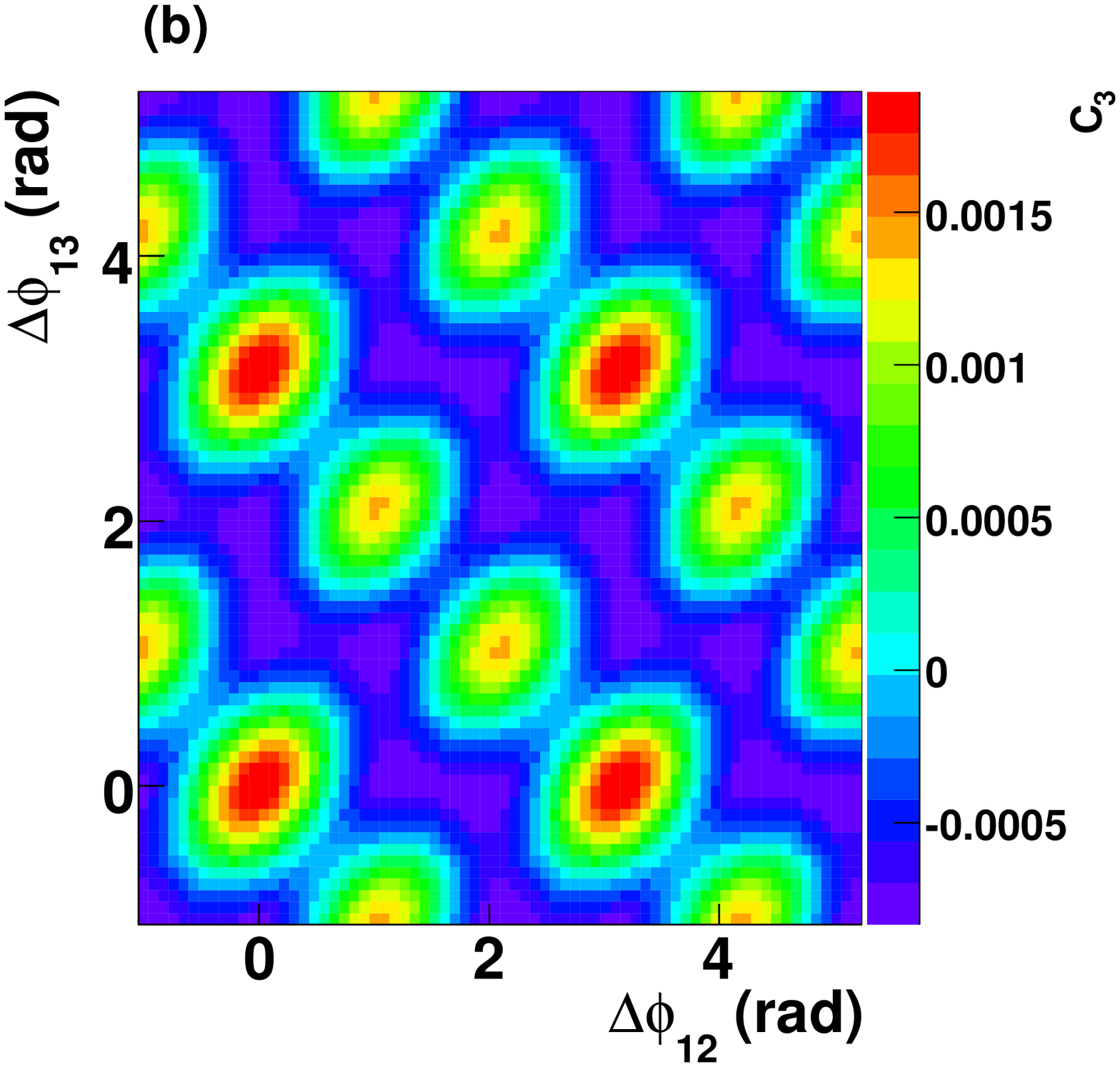}
  \end{center}
  \end{minipage}
  }
\mbox{
  \begin{minipage}{0.5\linewidth} 
  \begin{center}
  \includegraphics[width=0.99\linewidth]{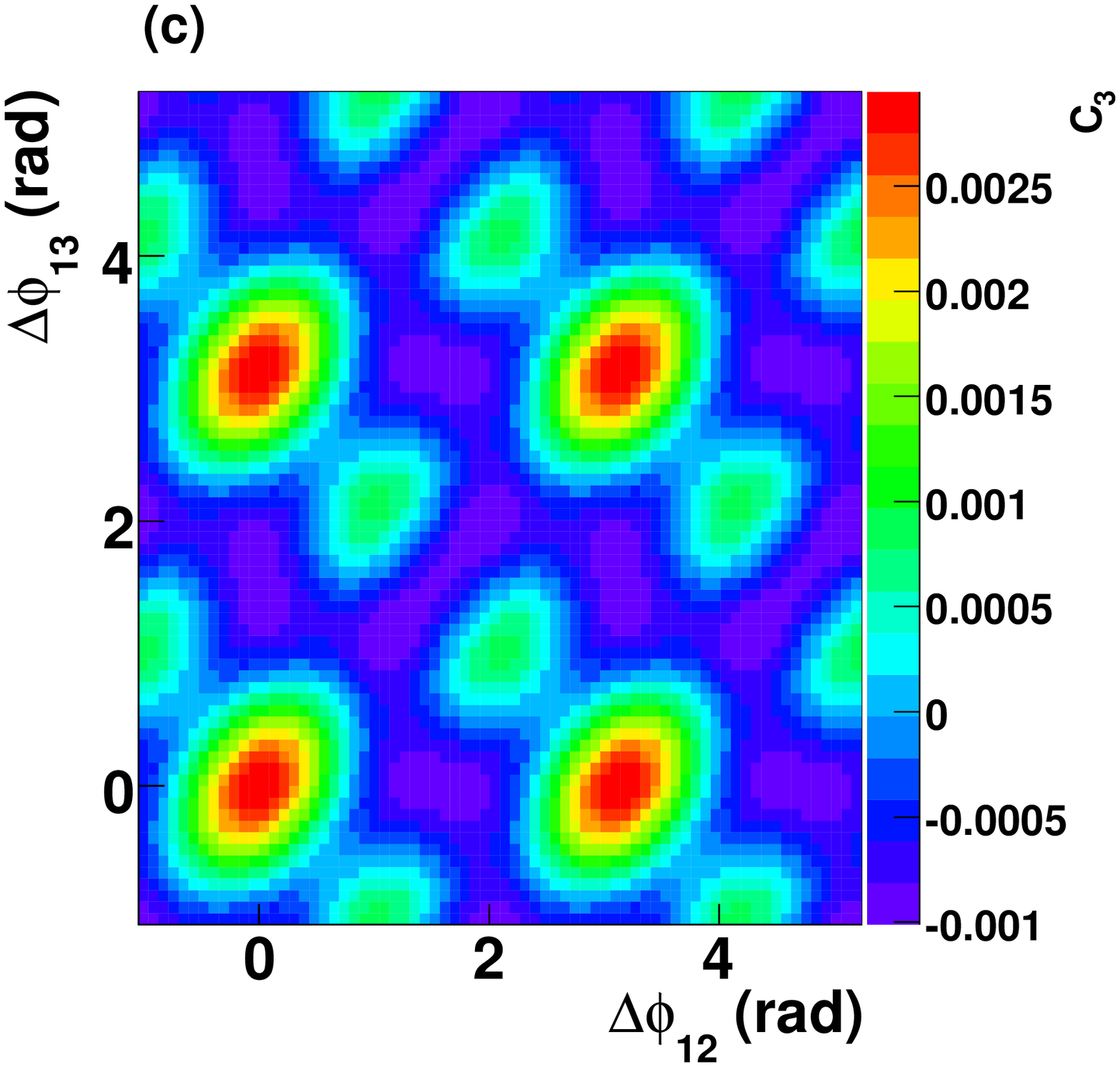}
  \end{center}
  \end{minipage}
  \begin{minipage}{0.5\linewidth}
  \begin{center}
  \includegraphics[width=0.99\linewidth]{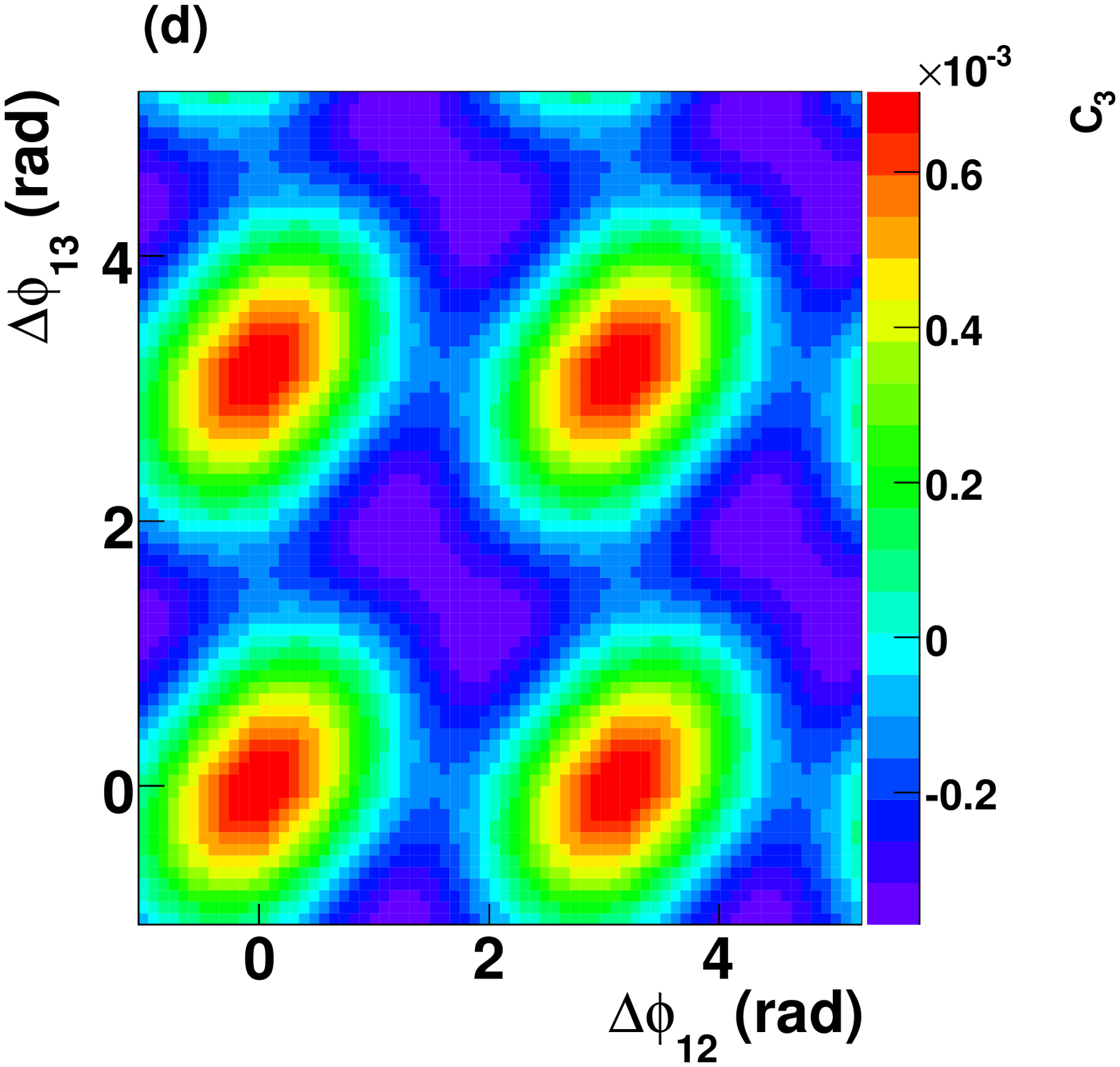}
\end{center}
  \end{minipage}
  }
\caption{3-cumulant flow signal calculated using (a) only  $v_2v_2v_4 $ terms, (b) 
 $v_2v_2v_4 $  and  $v_2v_2+ v_4v_4 $ terms in 1:0.5 ratio, (c)  $v_2v_2v_4 $   and  $v_2v_2+ v_4v_4 $ 
terms in a 1:2 ratio, (d) only  $v_2v_2+ v_4v_4 $ terms. The  $v_2v_2v_4 $   and  $v_2v_2+ v_4v_4 $ 
terms are calculated according to Eqs. \ref{c3Flow}-\ref{dfg}. See text for details.}
\label{fig2:flow}
\end{figure}

We find that the flow 3-cumulant involves, in general, an arbitrary combination of second (e.g.  $v_2^2 $) and third order (e.g.  $v_2v_2v_4 $) terms.
If particle production is strictly Poissonian, the coefficients  $f_{ijk} $ and  $g_{ijk} $ are unity. The second order terms of the cumulant thus vanish, and the 3-cumulant (flow only) 
then reduces to: 
\be
C_3^{F,Poisson} (\vphi_1,\vphi_2,\vphi_3 ) = \left( {2\pi } \right)^{ - 3} \lla {N_1 N_2 N_3 } \rra \Phi_3 (\vphi_1,\vphi_2,\vphi_3 ;v_p (1)v_m (2)v_n (3)) 
\label{c3FlowPoisson}
\ee
which features only non-diagonal terms  $v_m (i)v_n (j)v_p (k) $, i.e. with  $m \ne n $,  $m \ne p $, or  $n \ne p $.
This is illustrated in Figure \ref{fig2:flow} (a) which shows the flow 3-cumulant calculated using only  $v_2v_2v_4 $ terms. In 
general, particle production is non-Poissonian, and second order terms must be considered. The flow 3-cumulant shown in Fig. \ref{fig2:flow} (b) and (c) 
includes  $v_2^2 $ and  $v_4^2 $ added in ratios 1:0.5, and 1:2 respectively to the  $v_2v_2v_4 $ terms. 
Only  $v_2^2 $ and  $v_4^2 $ terms are included in Fig \ref{fig2:flow} (d)  illustrating a case where non-Poisson fluctuations are very large, and 
thereby dominate the 3-cumulant. The shape of the 3-cumulant thus depends significantly on the strength of the non-Poisson (2$^{nd}$ order terms), as well as
the magnitude of the flow coefficients. The interpretation of measured 3-cumulants may thus be considerably complicated by the presence of non-Poisson fluctuations.

Experimentally, one can estimate the magnitude of second order (diagonal) terms from 
measured total numbers of triplet, pair, and single particle densities. STAR observes based on data from Au + Au collisions at 200 GeV \cite{StarC308}, the 
coefficients  $f_{123} $ and  $f_{231} $ deviate from unity. The size of the deviation scales qualitatively as the inverse of the event multiplicity. STAR also finds the magnitude of the deviation depends on the  specific particle kinematic ranges used to carry the analysis. 
The particle production process is manifestly non-Poissonian.  Since the bulk of produced particles exhibits elliptic flow, this implies the conditions  $ f_{ijk}  = g_{ijk}  = 1$ are not verified in practice. 
The second order and constant terms of the cumulant do not vanish, and may in fact be sizable. The 2$^{nd}$ second order terms result from number fluctuations, and 
may in principle be estimated (and thus fitted) on the basis of measurements of $f_{ijk}$. Note however that multiplicity fluctuations occur for flow-like processes as well as all other types of production processes (including jets, conical emission, etc). It is thus non-trivial to 
unambiguously evaluate the proper magnitude of the diagonal flow terms based on ratios of measured yields. 

Effects of flow fluctuations are not specifically addressed in the above flow model. Note however that, as defined, the 3-cumulant measures averages of $\la v_n^2\ra$, 
and $\la v_i v_j v_k\ra$ coefficients rather than averages of $\la v_n \ra$. The 3-cumulant thus implicitely features fluctuations and non-flow effects. Given it is difficult to 
experimentally distinguish flow fluctuations and non-flow effects (see for instance the review by Voloshin {\it et al.} \cite{VoloshinReview08}), analyses attempting 
explicit subtraction of flow contributions (from three particle densities) based on the above equations, are thus intrinsically non-robust. 

\subsection{Conical Emission}
\label{Sect:ConicalEmission}
 
Mach cone emission of particles by partons propagating through dense QGP matter was proposed by 
Stoecker \cite{Stoecker05}  to explain the peculiar dip structure found at  $180^o $ in two-particle correlations 
reported by the STAR and PHENIX collaborations, and is the subject of many recent theoretical 
investigations \cite{Stoecker76,Stoecker05,JRuppert05,RupperMuller05,SolanaShuryak05,RenkRuppert06,Heinz06,Stocker07,RenkRupper07,MullerQM08,BBetz08,Armesto04,Salgado05a,RenkRuppert07_PLB646,Voloshin05,Hwa05_PRC74}.
The concept of Mach cone emission is based on the notion that 
high momentum partons propagating through a dense QGP interact with the medium and loose energy 
(and momentum) at a finite rate. The release of energy engenders a wake that propagates at a 
characteristic angle, the Mach angle, determined by the sound velocity in the medium.  Authors of Ref.   
\cite{SolanaShuryak05} estimated the speed of sound in the QGP to be of the order of  $v_s \approx c_s^{RHIC} \approx 0.33 $.
The Mach angle should thus be of the order of  $70^o $ relative to the away-side parton 
direction. We use this prediction to motivate a simple geometrical model of conical emission. The near-
side jet is described using a Gaussian azimuthal profile. Away side particles are emitted at a cone angle of 
1 radian with a scatter of 0.05 radian relative to the away-side direction. Figures \ref{Fig:Cone1} (a-b)  present 
the 3-cumulant obtained with a Monte Carlo simulation based on two million 
events. Given that Mach cone particles are produced at 1 radian  from the away-side direction and 
roughly normal to the beam direction, narrow Jacobian peaks are seen in the three-particle correlations.  
A strong dip is present at  $180^o $ in the two-particle correlations, while in the three-particle 
correlation a clear spacing is found between the peaks. In this 
simple model,the finite width of the peaks is due in part to the finite width of the trigger jet, and in part to the scatter imparted to the cone 
particles relative to the away-side direction. In practice, one might expect additional broadening of the 
cone because the speed of sound changes through the life of the QGP medium, and given the finite size 
of the medium.  Radial flow might also significantly alter the correlation functions. As discussed in 
\cite{Pruneau06}, the details of the correlation shapes clearly depend on assumptions made about the 
kinematics of the away-side parton. 

\subsection{Flow  $\times $ Jet Correlations}
\label{FlowJetCorrelations}

The measured nuclear modification factor,  $R_{AA} $, defined as the ratio single particle yields measured in  $A+A $ collision to those 
measured in  $p+p $ interactions (scaled by the number of binary collisions in  $A+A $) suggest the 
propagation of jets through the medium is severely quenched, and therefore features a surface emission 
bias \cite{WhitePapers}. In this scenario, partons propagating through the medium loose a large 
fraction of their energy. This results into jet lost or lower energy jets with smaller particle multiplicity. 
The path length through the medium determines the amount of energy loss and quenching. At a given 
point of emission, near the medium surface, a parton propagating outward in a direction normal to the 
surface should have shortest pathway through the medium, and minimal energy loss. Partons produced 
at the same location, but emitted in other directions would have longer path through the medium and 
suffer larger energy loss. In this scenario, one thus expects jet yields (and consequently the high  $p_t $
 trigger) to be correlated to the reaction plane orientation. Neglecting disturbances imparted to the 
medium by the propagation of the jet (or parton), we model the jet dependency on azimuthal angle 
relative to the reaction plane with Fourier decomposition. Specifically, we write the probability of the 
jet being emitted at angle  $\phi$ 
 while the reaction plane is at angle  $\psi$ as	
\begin{equation}
P(\phi |\psi ) = 1 + 2\sum\limits_n {a_n \cos } \left( {n(\phi  - \psi )} \right)
\label{diffAttn}
\end{equation}	
where the coefficients  $a_n $ represent the effect of the differential azimuthal attenuation. We 
parameterize the jet multiplicity and azimuthal width using associated yields, and Gaussian widths that 
do not depend on the azimuthal direction. As in Sec. \ref{Sect:DiJetProduction}, the average number of jets per 
event is written  $\lla J \rra  $
 while the number of particles associated with a given jet is $\lla {A_i } \rra $.
We also assume the presence of a {\it flowing} background consisting on average of   $\lla {B_i} \rra  $
particles.  The single particle yield is thus  $\rho_{1,J \otimes F} (\vphii ) = (2\pi )^{ - 1} \left( {\lla J \rra \lla A_i\rra  + \lla B_i\rra}\right) $. 
Given the production of both jets, and background particles is correlated to the reaction plane, one 
ends up with flow-induced correlations between all particles, even those produced by the jets. The two-particle 
cumulant involves three sets of terms, namely jet-jet ($J \otimes J$), jet-background ($J\otimes B$), and background-background 
($B \otimes B$). Similarly, the three-particle cumulant involves $J \otimes J \otimes J$,
  $J \otimes J \otimes B $,  $J \otimes B \otimes B $, and   $B \otimes B \otimes B $ terms. 

A jet-flow model was already discussed in \cite{Pruneau06}. We here re-derive, and express the 
various terms of the model with a more compact and intuitive notation. The terms,  $B \otimes B $
and  $B \otimes B \otimes B $
 are identical in form to those considered in Sect \ref{Sect:flow}. We need to discuss only the terms $J \otimes J$,
 $J \otimes B $,  $J \otimes J \otimes J $,  $J \otimes J \otimes B $, and $J \otimes B \otimes B $.
 We begin with $J \otimes J $, and $J \otimes J \otimes J $.  
These terms contain contributions where two (three) of the particles are part of the same jet, and 
others where one, or two particles are not from the same jet. When particles are from the same jet, the 
correlation to the reaction plane is unimportant (to the extent, in our model, that the width of the jets do 
not change with their orientation relative to the reaction plane). The corresponding terms are thus 
identical to those obtained in the absence of flow.  The two-particle density includes a term with two 
particles from the same jet (identical to that discussed in Sec. \ref{Sect:DiJetProduction}), and a term with particles from 
two different jets. This last term is given by:
\bea
\lla {J(J - 1)} \rra \lla {A_i } \rra \lla {A_j } \rra \int{d\psi d\phi_\alpha  G(\vphii  - \phi_\alpha  )}\\
\times P(\phi_\alpha  |\psi )\int {d\phi_\beta  G(\vphii- \phi_\beta  )P(\phi_\beta  |\psi )} \nonumber
\eea
Integration, and inclusion of the first term yields the jet component of the two-particle density.
\bea
\rho_2^{Jet only} (\vphii,\vphij ) &=& \lla J \rra \lla {A_i A_j } \rra G_2(\vphii  - \vphij ;\sigma_{12} ) \\ 
  & &+\lla {J(J - 1)} \rra \lla {A_i } \rra \lla {A_j } \rra F_2(\vphii  - \vphij ;a_n (i,j)) \nonumber
\eea
where  $\sigma_{12}^2=\sigma_1^2  + \sigma_2^2  $,  $a_n (i,j) = v_n '(i)v_n '(j) $, with $v_n '(i) = a_n \exp ( - n^2 \si^2 /2) $.
Similarly, the three-particle density includes a term with three particle from the same jet, two particles 
from the same jet, and another with all three particles from different jets.  The $J \otimes J $, and $J \otimes J \otimes J $ part og the cumulant is as follows.
\bea
 \rho_3^{Jet only} (\vphii,\vphij,\vphik ) &=& \lla J \rra \lla {A_i A_j A_k } \rra G_3 (\vphii,\vphij,\vphik ;\si,\sj,\sk ) \\ 
& &+\lla J(J - 1)\rra \lla A_i A_j\rra \lla A_k\rra G_2(\Delta\vphi_{ij})F_2\left({\frac{\sj^2 }{\sij^2}\vphii+\frac{\si^2}{\sij^2}\vphij-\vphik;v_n''(k)}\right)+(jki)\nonumber \\
& &+\lla J(J - 1)(J - 2)\rra \lla A_i\rra \lla A_j\rra \lla A_k\rra N_3(\vphii,\vphij,\vphik;v_n'(i),v_m'(j),v_p'(k)) \nonumber
\label{JJJ}
\eea
The functions  $G_3 $,  $F_3 $,  $F_2 $ were defined in previous sections.  The symbol  $(jki) $ indicates 
identical terms obtained by permutation of the indices. The term $G_3 $  is identical to that obtained for 
non-flowing jets. The term of the last row corresponds to a flow-like correlation between the three 
measured particles, albeit with a strength which depends on the fluctuations of the number of jets per 
event, and the jet associated multiplicities in the kinematic ranges considered.  The second row contains 
three similar terms that embody the jet-flow correlations: two particles are correlated because they 
belong to the same jet, while the third is correlated to the first two because of reaction plane 
dependencies. One notes that  in the above expression, $F_2 $ depends on the angle of emission of all three particles (at variance 
with assumptions made in Ref. \cite{Ulery06a,Ulery06b})  in a non trivial way. Indeed, in 
general, the jet correlation widths  $\si $ depend on the particle momentum ranges: the ratios 
 ${{\si^2 } \mathord{\left/  {\vphantom {{\si^2 } {\sij^2 }}} \right. \kern-\nulldelimiterspace} {\sij^2 }} $
are arbitrary (non-integer) values.  $F_2 $ therefore involves inharmonic flow components. This 
implies it is inappropriate to model jet-flow cross-terms as a simple product of jet-like correlations, and 
flow-like correlations, as in Ref. \cite{Ulery06a,Ulery06b}, for purposes of background subtraction.

We next consider the component  $J \otimes J \otimes B $, and  $J \otimes B \otimes B $ of the 3-cumulant. 
Following the steps used for the derivation of the pure jet components, one finds it also includes a term 
proportional to  $F_2 G_2 $ indistinguishable from  $J \otimes J \otimes J$. One has
\be
\lla J \rra \lla {A_1 A_2 } \rra \lla {B_3 } \rra G_2(\Delta\vphi_{12};\sigma_{12} )F\left( {\frac{{\sigma_2^2 }}{{\sigma_{12}^2 }}\vphi_1  
+ \frac{{\sigma_1^2 }}{{\sigma_{12}^2 }}\vphi_2  - \vphi_3 ;v_n ''(3)} \right)
\label{JJB}
\ee
The  $J \otimes B \otimes B $ compoment of the 3-cumulant contains a non-diagonal flow term as follows
\be
\lla J \rra \lla {A_1 } \rra \lla {B_2 B_3 } \rra F_3 (\vphii,\vphij,\vphik ;v_n '(i),v_m '(j),v_p '(k))
\label{JBB}
\ee
This term has the same functional dependence and is indisguishable from the term on the third line of Eq \ref{JJJ}.
Assemblinging all components, one obtains the JET  $ \otimes  $
FLOW 3-cumulant:

\bea
 C_3^{J \otimes F} (\vphii,\vphij,\vphik ) &=& \lla J \rra \lla {A_i A_j A_k } \rra G_3 (\vphii,\vphij,\vphik ;\si,\sj,\sigma _k ) \nonumber \\ 
  &+& \left( {\lla {J(J - 1)} \rra  - \lla J \rra ^2 } \right)\lla {A_i A_j } \rra \lla {A_k } \rra G_2(\Delta\vphi_{ij}) + (jki)  \nonumber\\ 
  &+& \lla {A_i A_j } \rra \left( {\lla {J(J - 1)} \rra \lla {A_k } \rra  + \lla J \rra \lla {B_k }\rra}\right)G(\Delta\vphi_{ij};\sij)\Phi_2\left({\frac{{\sj^2 }}{{\sij^2 }}\vphii-\frac{{\si^2 }}{{\sij^2 }}\vphij  - \vphik ;v_n ''(k)} \right) + (jki) \\ 
  &-& \lla {J(J - 1)} \rra \lla {A_i } \rra \lla {A_j } \rra \lla {A_k } \rra F_2 (\vphii  - \vphij ;a_n (i,j)) + (jki) \nonumber\\ 
  &+& \lla {J(J - 1)(J - 2)} \rra \lla {A_i } \rra \lla {A_j } \rra \lla {A_k } \rra N_3 (\vphii,\vphij,\vphik ;v_n'(i),v_m'(j),v_p '(k) \nonumber\\ 
  &+& \lla J \rra \lla {A_i } \rra \lla {B_j B_k } \rra N_3 (\vphii,\vphij,\vphik ;v_n '(i),v_m '(j),v_p '(k)) \nonumber\\ 
  &-& \lla {B_i B_j } \rra \lla J \rra \lla {A_k } \rra F_2 \left( {\vphii  - \vphij ;v_n (i)v_n (j)} \right) + (jki)  \nonumber\\ 
  &+& \left( {2\pi } \right)^{ - 3} \lla {B_i B_j B_k } \rra \Phi_3 (\vphii,\vphij,\vphik ;v_p (i)v_m (j)v_n (k)_k ) \nonumber\\ 
  &+& \left( {2\pi } \right)^{ - 3} \lla {B_i B_j B_k } \rra \left( {1 - f_{ijk} } \right)\Phi_2 (\vphii  - \vphij ;v_m (i),v_m (j)) + (jki)  \nonumber\\ 
  &-& \left( {2\pi } \right)^{ - 3} \lla {B_i B_j B_k } \rra f_{ijk}  + (jki)  \nonumber \\
  &+& constants 
  \label{JetFlowCumulant}
\eea
This cumulant includes components typical of jets (line 1 and 2), flow (line 4-11'), and one term 
unique to jet-flow cross-term (line 3).  While this jet-flow term does include two-particle jet-like, and flow-like factors, 
it is important to note the flow factor is intrinsically inharmonic. Its amplitude depends on the number 
jet associates and background particle multiplicity. Given the number of associates is likely much 
smaller than the number of background particles, one expects the amplitude of this term to be 
dominated by the magnitude of the background.  It is interesting to compare the magnitude of this 
cross-term relative to the off-diagonal flow terms. We focus on the difference between the leading 
terms in  $v_2v_2v_4 $  and $v_2v_2$.  Given the jet cross section and fragment multiplicity are small at 
RHIC energies, one expects the   $v_2v_2v_4 $  term should dominate or at most be of similar 
magnitude to the cross-term unless the jet flow  $v_2 $ is considerably larger than the bulk flow. 

As discussed in Sect \ref{Sect:flow}, particle emission in  $A+A $ collisions is a non-Poissonian process. The coefficients
$f_{ijk}$ in general deviate from unity. The leading background terms may thus be those of line 2, 6, 7, and 9 in Eq. \ref{JetFlowCumulant}.

Lines 4 and 7 both contain terms in $F_2$ albeit with different amplitudes. Experimentally these cannot be distinguished, and one ends up 
with an $F_2$ contribution to the cumulant which depends on both the jet and background yields.  Likewise, lines 5, 6, 8, and 9 feature non-diagonal 
flow terms (dominated by $v_2v_2v_4$) which cannot be distinguished experimentally: the amplitude of the $v_2v_2v_4$ three particle correlation terms 
depend intricately on the jet yield, its fluctuations, as well as the background yield.  The jet-flow 'cross-term' manifests itself through
the inharmonic terms of line 3 in Eq. \ref{JetFlowCumulant}. While these harmonic terms may be approximated as a 
product such as $G_2(\Delta\phi_{12})F\Phi_2(\Delta\phi_{13})$ if the width $\sigma_1$  (very high-$p_t$ particles) is negligible, when
added in quadrature, to the width $\sigma_2$ of the low-$p_t$ particle, the approximation breaks down in general 
because $\sigma_1$ is neither zero or equal to $\sigma_2$. Additionally, note that the model used in this section assumed 
the cross-terms are dominated by jetty and flow components. This may not be the case in practice.
For instance resonance or cluster decays in the presence of both radial and elliptical flow should lead to complex cross terms. Such cross-terms 
shall be inhenrently inharmonic also. Given the correlation shapes associated with resonance or cluster decays can be quite wide, the inharmonic 
character of the cross term can be quite intricate. Adhoc subtraction of terms in $G_2(\Delta\phi_{12})F\Phi_2(\Delta\phi_{13})$  in model based 
analyses is therefore unwarranted.

\section{Probability Cumulants}
\label{Sect:ProbabilityCumulants}

We showed in Section \ref{Sect:flow} that the 3-cumulant associated with flow reduces to non-
diagonal terms in  $v_n v_m v_p  $ for Poissonian particle production processes. However, momentum, energy, and quantum number 
conservation laws imply elementary collisions are typically non-Poissonian processes as observed from 
Fig 3 which shows ratios  $f_{ijk}$ measured by STAR \cite{StarC308}. This implies the 3-cumulant may have a rather complicated 
structure, with second order terms (i.e. two-particles) as well as three particle terms.  We note however 
that the simplicity of the 3-cumulant may be recovered, in this case, by using  probability cumulants 
rather than density cumulants. The probability cumulants are defined as follows:
\bea
 P_2 (\vphii,\vphij)         &\equiv& \frac{{\rho_2 (\vphii,\vphij )}}{{\lla {N_i N_j } \rra }} - \frac{{\rho_1 (\vphii )\rho_1 (\vphij )}}{{\lla {N_i } \rra \lla {N_j } \rra }} \\ 
 P_3 (\vphii,\vphij,\vphik ) &\equiv& \frac{{\rho_3 (\vphii,\vphij,\vphik )}}{{\lla {N_i N_j N_k } \rra }}- \frac{{\rho_2 (\vphii,\vphij )\rho_1(\vphik )}}{{\lla {N_i N_j } \rra \lla {N_k } \rra }} \nonumber\\
  &-& \frac{{\rho_2 (\vphii,\vphik )\rho_1 (\vphij )}}{{\lla {N_i N_k } \rra \lla {N_j } \rra }} - \frac{{\rho_2 (\vphij,\vphik )\rho_1 (\vphii )}}{{\lla {N_j N_k } \rra \lla {N_i } \rra }} \nonumber \\ 
  &+& 2\frac{{\rho_1 (\vphii )\rho_1 (\vphij )\rho_1 (\vphik )}}{{\lla {N_i } \rra \lla {N_j } \rra \lla {N_k } \rra }} \nonumber \\ 
\label{probabilityCumulant}
\eea	
where  $N_i $ correspond to total particle multiplicities accepted in the kinematic cuts  $i $.  It is 
straightforward to verify that the flow probability 3-cumulant indeed consists of non-diagonal flow terms only:
\be
P^N_3(\vphi_1,\vphi_2,\vphi_3 ) = \left( {2\pi } \right)^{ - 3} \Phi_3 (\vphi_1,\vphi_2, \vphi_3 ;v_p (1)v_m (2)v_n (3))
\label{flowProbabilityCumulant}
\ee
The probability cumulant,  $P_3$, thus provides a tool to suppress the strength of non-Poissonian 
2$^{nd} $ order terms, and may therefore be used, in addition to the number 3-cumulant,   $C_3$, in three-particle analyses. 

The simplification obtained for flow processes is unfortunately not realized for jet or conical emission processes. The use of probability 3-cumulant
may then be of limited interest in practice. This can be straightforwardly shown through a calculation of the probability cumulant of the Gaussian 
model used in Sec. \ref{Sect:DiJetProduction}. We limit our calculation to the near side jet. One finds the probability 3-cumulant
contains terms  in $G_2(\Delta\phi)$ proportional to the non vanishing factor 
$\lla {J(J - 1)} \rra \lla {N_1 N_2 } \rra \lla {N_3 } \rra  - \lla J \rra ^2 \lla {N_1 N_2 N_3 } \rra $, where ${N_1}$, ${N_1 N_2 }$, and ${N_1 N_2 N_3}$
are respectively the number of singles, pairs, and triplets from the near side jet. By contrast to the flow probability 3-cumulant which contains
only genuine three-particle correlation terms ($\Phi_3$), the Gaussian probability cumulant thus has a complicated expression which contains terms 
in $G_3$ as well as in $G_2$. Its interpretation is therefore non trivial.

\section{Efficiency Correction and Observable Robustness}
\label{Sect:EfficiencyCorrection}

Measured particle densities, and cumulants must be corrected for detection efficiencies and other 
instrumental effects. We introduced in Ref. \cite{Pruneau06} a procedure to correct for detector 
efficiencies based on ratios of two- and three-particle densities by products of two, and three single 
particle densities. For instance, up to a global efficiency factor, the corrected three-particle density is 
given by
\be
\rho_3(\vphi_1,\vphi_2,\vphi_3 )_{corrected}  = \left( {2\pi } \right)^{ - 3} N(1)N(2)N(3)\frac{{N_3 (\vphi_1,\vphi_2,\vphi_3 )}}{{N_1 (\vphi_1 )N_1 (\vphi_2 
)N_1 (\vphi_3 )}}
\label{correction}
\ee
where  $N_1(\vphii) $ stands for uncorrected single particle densities,  $N_3 (\vphi_1,\vphi_2,\vphi_3 ) $
are  numbers of triplets, and  $N(i) $ are total particle yields within the kinematics cuts $i$.

This correction procedure is strictly exact for continuous functions provided the efficiency is a function 
of the azimuthal angles  $\vphii  $ but independent of other observables such as the particle 
rapidity and transverse momentum. For large 
detectors such as the STAR TPC \cite{StarNim}, the efficiency is a smooth and slowly varying function of the 
pseudorapidity and transverse momentum, but exhibit periodic structures in azimuth because of TPC 
sector boundaries. A correction for azimuthal dependencies of the detector response is thus the most important and relevant 
in this context. 

\begin{figure}[!htP]
\includegraphics[width=3.5in]{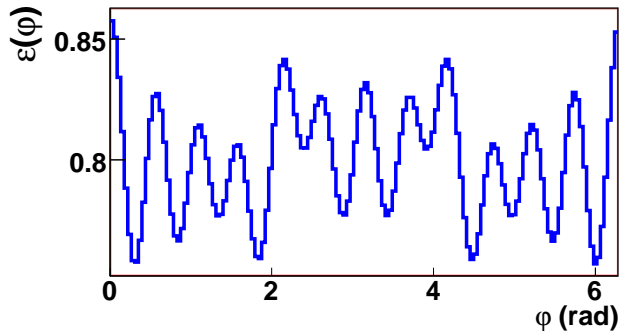}
\caption{(Color online) Detection efficiency dependence on the particle azimuthal angle used in the 
simulation results presented in Fig. \ref{Fig:EfficiencyStudy}.}
\label{fig4:EfficiencyVsPhi}
\end{figure}

\begin{figure}[!htP]
\mbox{
  \begin{minipage}{0.5\linewidth} 
  \begin{center}
  \includegraphics[width=0.99\linewidth]{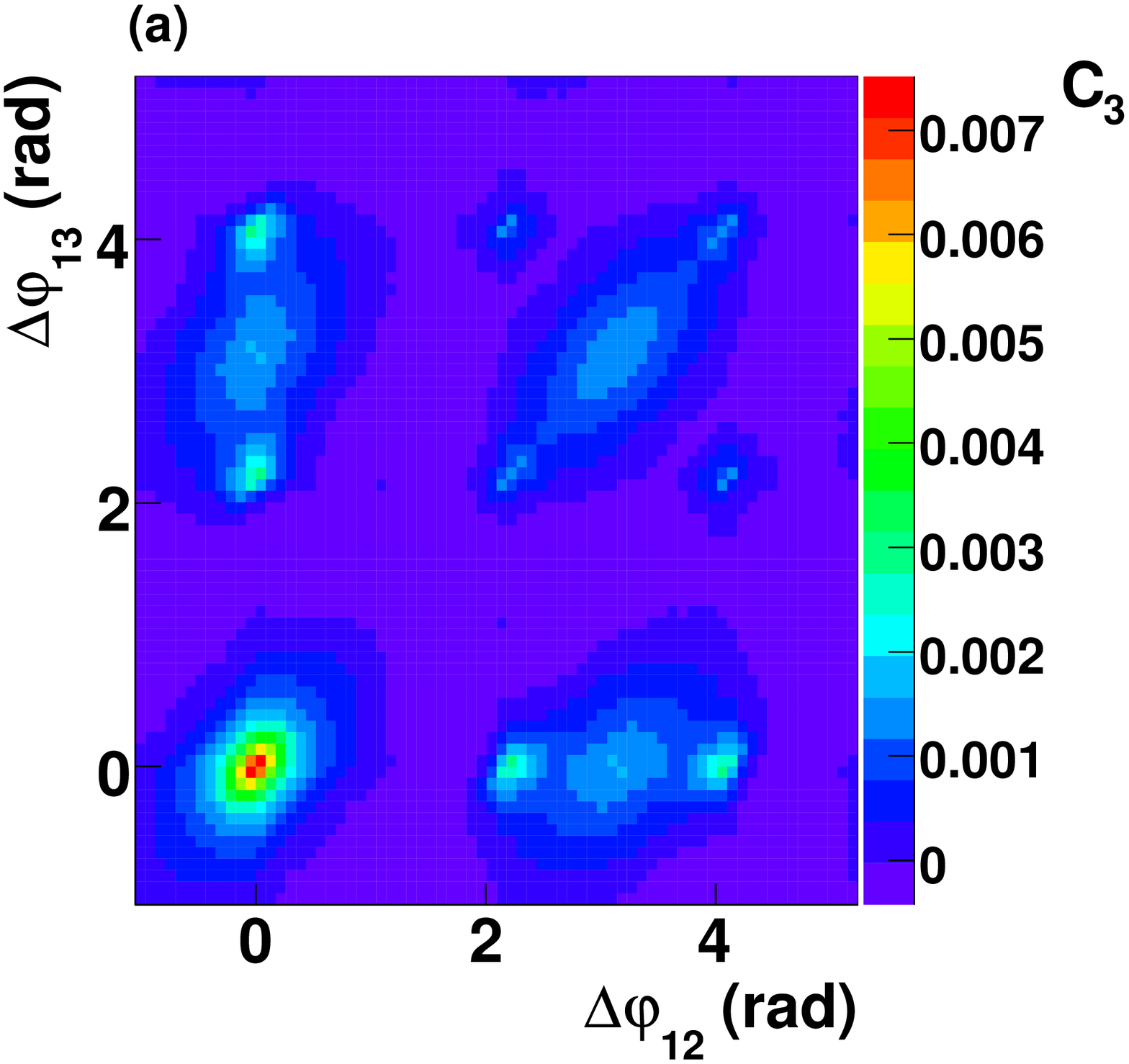}
  \end{center}
  \end{minipage}
  \begin{minipage}{0.5\linewidth} 
  \begin{center}
  \includegraphics[width=0.99\linewidth]{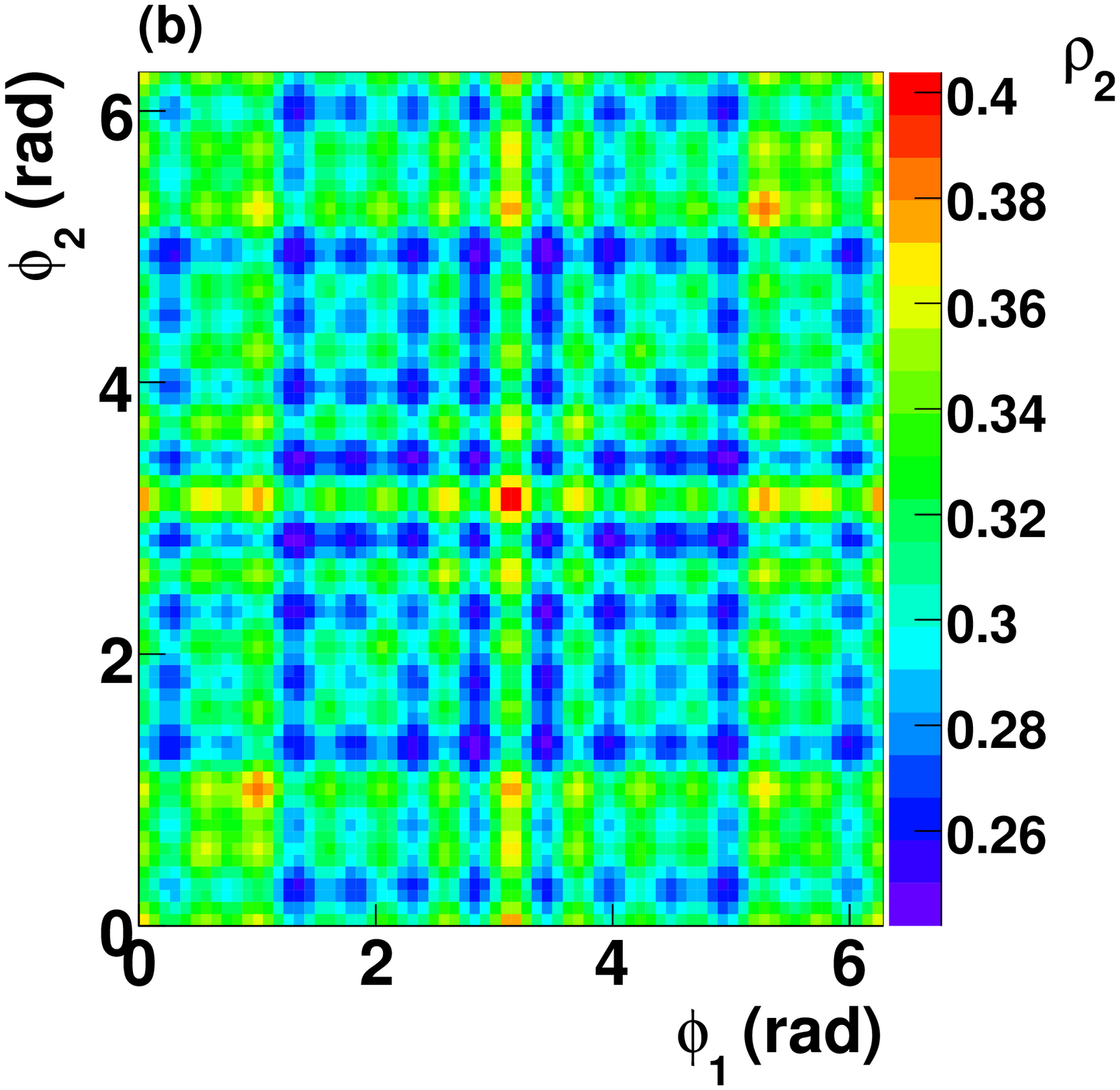}
  \end{center}
  \end{minipage}
  }
\mbox{
  \begin{minipage}{0.5\linewidth} 
  \begin{center}
  \includegraphics[width=0.99\linewidth]{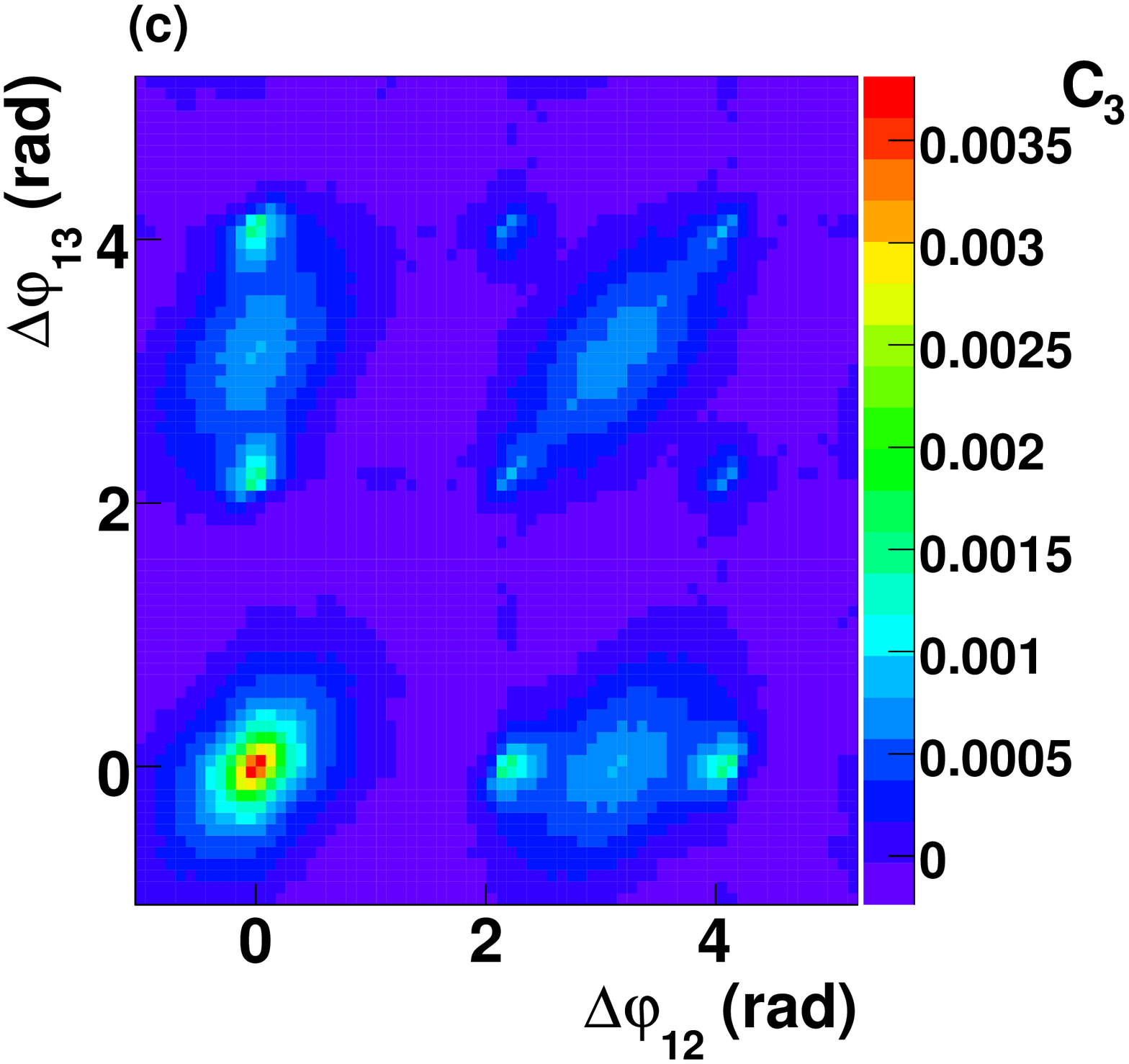}
  \end{center}
  \end{minipage}
  \begin{minipage}{0.5\linewidth}\begin{center}
  \includegraphics[width=0.99\linewidth]{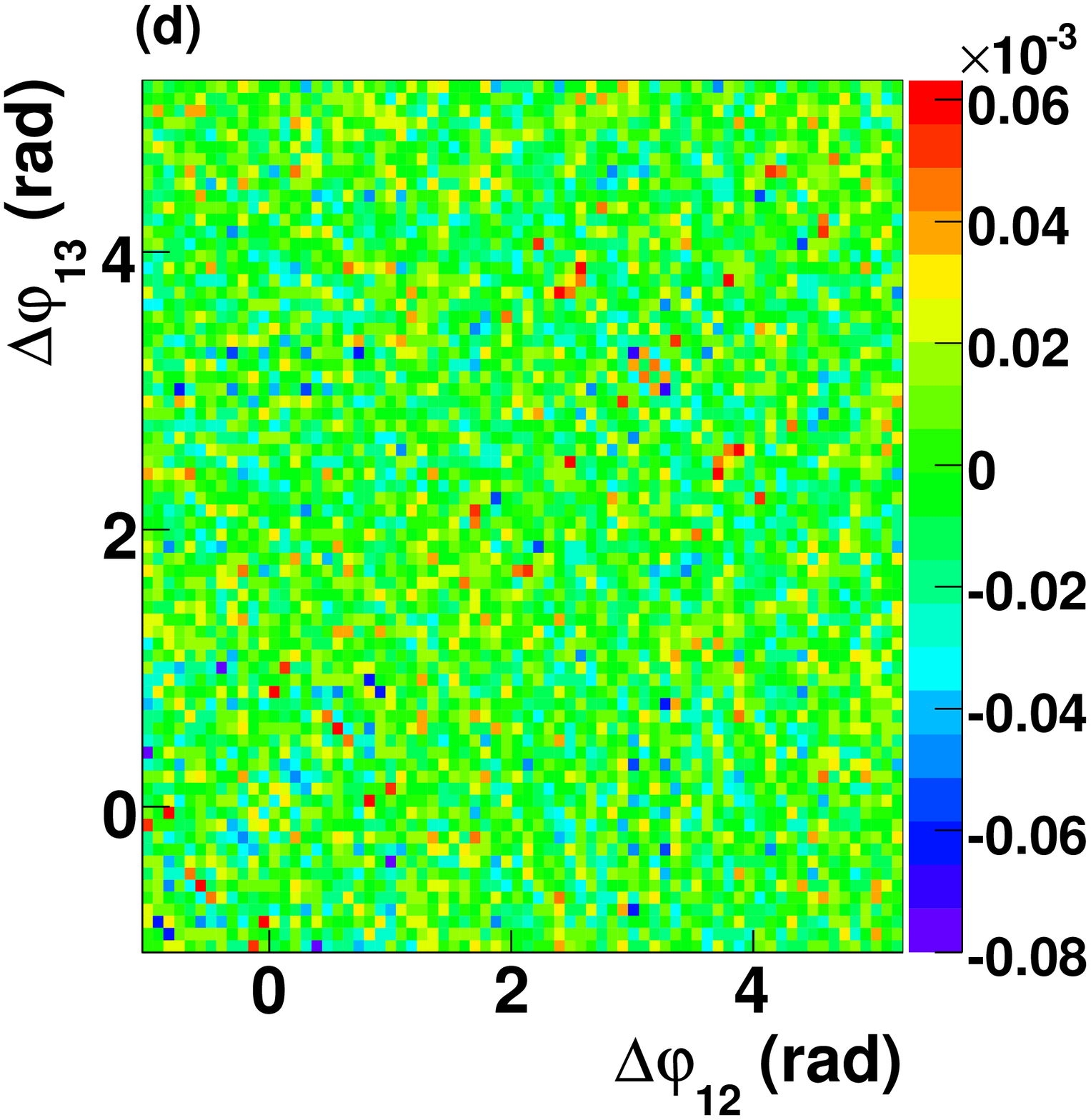}
  \end{center}
  \end{minipage}
  }
\caption{Study of the efficiency correction method (Eq. \ref{correction}) using the Mach Cone 
model described in Sec. \ref{Sect:Sensitivity}. (a) 3-cumulant for perfect detection efficiency, (b)
Two particle density, (c) 3-cumulant obtained with the azimuthal efficiency  dependence shown in Fig. \ref{fig4:EfficiencyVsPhi}, (d) Difference 
between the 3-cumulants shown in (b), and the perfect efficiency cumulant scaled by the cube of the 
average detection efficiency. }
\label{Fig:EfficiencyStudy}
\end{figure}

We examine the robustness of the above correction procedure in a practical situation, i.e.  where the 
densities are measured with a finite number of bins,  based on the jet and conical emission models 
discussed in Sec. 2. We parameterize the detection efficiency,  $\varepsilon (\vphii ) $, with a Fourier decomposition.
\be
\varepsilon (\vphii ) = \varepsilon_0 \left( {1 + 2\sum\limits_{n = 1}^\infty  {\varepsilon_i \cos 
\left( {n\vphii } \right)} } \right)
\label{response}
\ee
where  $\varepsilon_0 $ is the average efficiency, and coefficients  $\varepsilon_n  $
determine the azimuthal dependence of the efficiency. We assume the efficiency for simultaneously 
detecting two and three particles are factorizable as product of single particle efficiencies. The mean 
single particle detection efficiency is set to 80\%.  The Fourier coefficients (Eq. \ref{response}) are chosen to 
obtain a non-uniform azimuthal detector response as shown in Fig. \ref{fig4:EfficiencyVsPhi}. 
The JET+Mach Cone model described in sections \ref{Sect:ConicalEmission} and \ref{Sect:Sensitivity} is used to carry a simulation of the robustness of 
the correction procedure. Figure \ref{Fig:EfficiencyStudy} (a) displays the 3-cumulants obtain with perfect detection
efficiency.  Figure \ref{Fig:EfficiencyStudy} (b) shows the uncorrected two particle density $\rho_2(\vphi_1,\vphi_2)$
obtained with the non-uniform azimuthal response illustrated in Fig \ref{fig4:EfficiencyVsPhi}. The two particle density exhibits
strong, narrow, and repeated structures that result from the twelve-fold structure of the assumed detector response.
Figure \ref{Fig:EfficiencyStudy} (c) displays the 3-cumulant obtained with the same detector response, but corrected 
for efficiency effects by division of the two and three particle densities by the product of single particle densities, as in Eq. \ref{correction}.
In stark contrast to the correlation shown in Figure \ref{Fig:EfficiencyStudy} (b), one finds the corrected cumulant exhibits no evidence of the detector response except for a global 
change in amplitude corresponding to the cube of the average detection efficiency.  Figure \ref{Fig:EfficiencyStudy} (d)
shows the difference between this cumulant and the perfect efficiency cumulant scaled down by the cube of the average 
detection efficiency. The standard deviation of the difference is smaller than 1\% of the maximum 3-cumulant amplitude
and exhibit no particular structure.  We thus conclude the correction method is numerically robust for cases 
where the triplet, and pair efficiencies factorize.

\section{Cumulant Method Sensitivity For Jet and Conical Emission Measurements}
\label{Sect:Sensitivity}

We present in this section a study of the sensitivity of the cumulant method described in this paper for 
a search for conical emission. Our study of the sensitivity of the method is based on 
Monte Carlo (MC) simulations carried out using a simple event generator that encapsulates flow, jet, 
and conical emission as a multicomponent model. 

We begin our presentation of the simulations with flow-only model components. The overall 
multiplicity, noted $m$, of an event is selected randomly according to a flat distribution.  The flow component is 
designed to produce, on average, a fraction,  $f_i$, of the event multiplicity. The number of {\it flowing} particles,  $N_i$,
 (for a kinematic cut $i$) is generated event-by-event randomly with a Poisson probability density 
function (PDF) of mean  $m \times f_i $.  The particle direction, in azimuth, is selected randomly according to the flow PDF given by Eq. 
\ref{flowFourier}. Various values of the coefficients  $v_2$ and  $v_4$, which  determine the magnitude of the bulk flow,  
are used in the following to study the impact of flow on the cumulant and sensitivity to a cone signal.  
All other Fourier coefficients are set to zero. The event plane azimuthal angle is chosen 
randomly in the range  $[0, 2\pi[$.  The coefficient  $f_1$ is set to produce a number of high-$p_t$ 
particles or order unity in each event, while $f_2$ is set to produce a low  $p_t$ particle multiplicity of 
order 100. These values are selected to mimic STAR data in the range 
 $3<p_t<20 $, and  $1<p_t<2$ for the trigger and associate particles respectively.

\begin{figure}[!htP]
\mbox{
  \begin{minipage}{0.33\linewidth} 
  \begin{center}
  \includegraphics[width=0.99\linewidth]{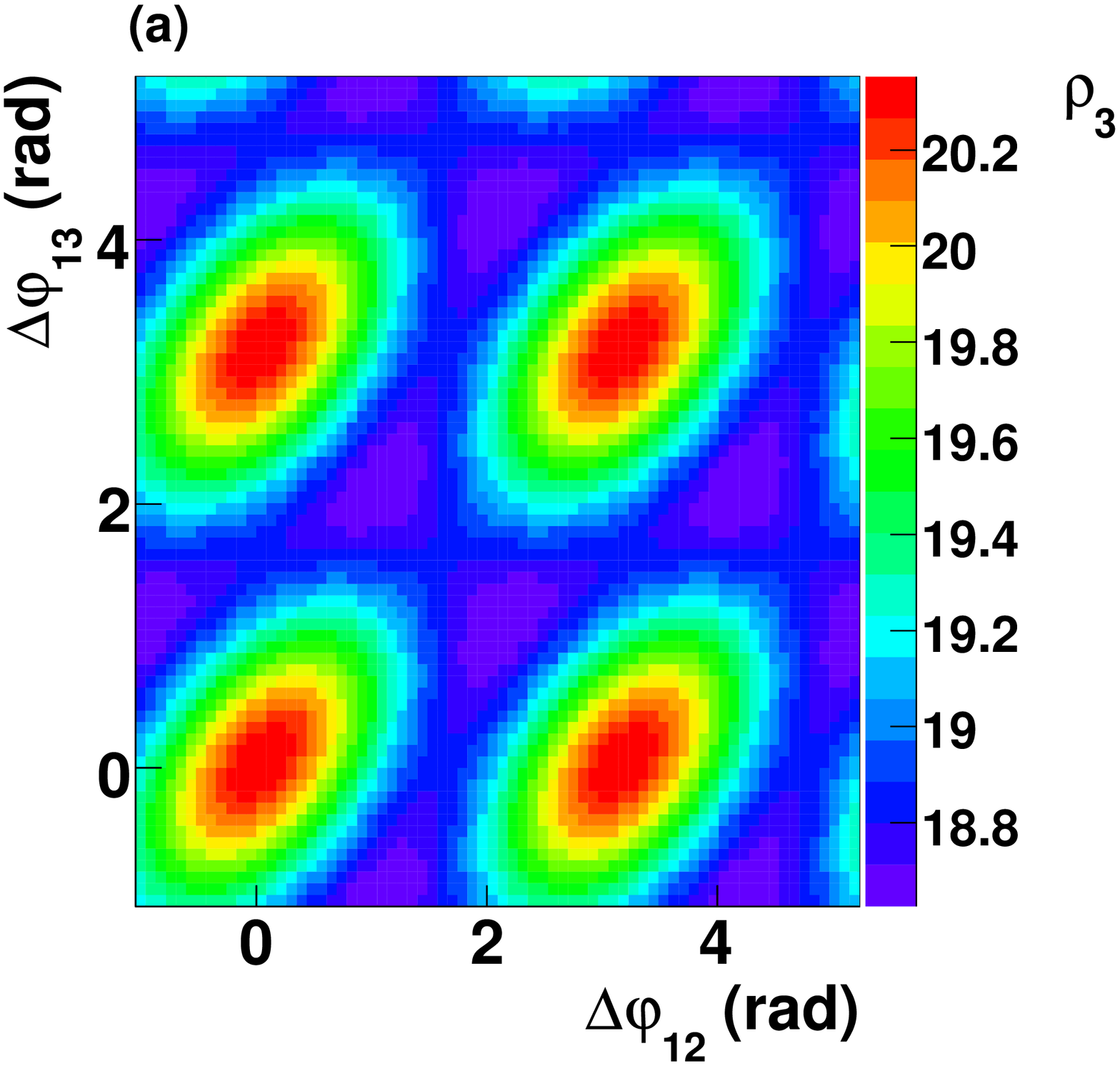}
  \end{center}
  \end{minipage}
  \begin{minipage}{0.33\linewidth} 
  \begin{center}
  \includegraphics[width=0.99\linewidth]{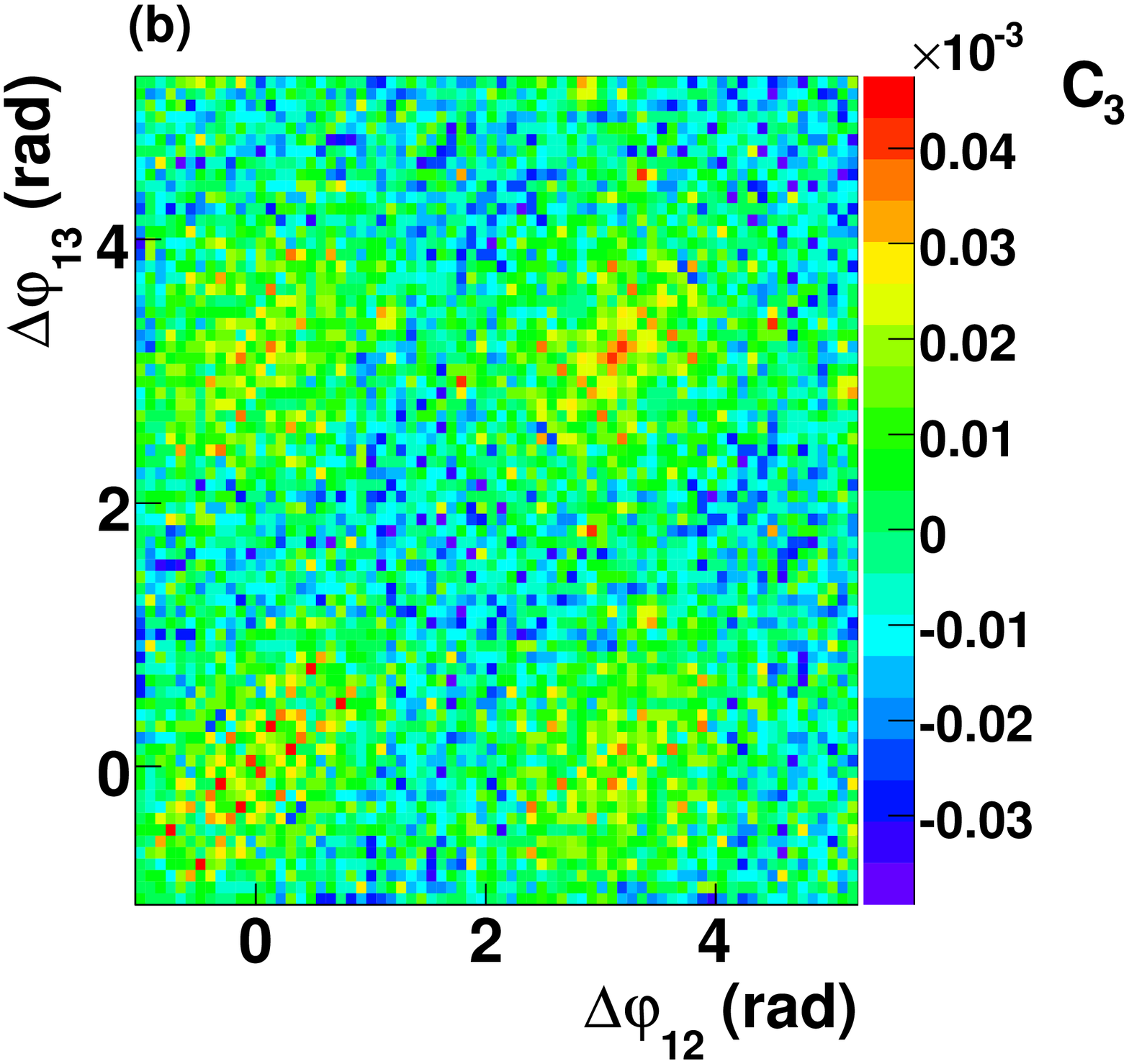} 
  \end{center}
  \end{minipage}
  \begin{minipage}{0.33\linewidth}\begin{center}
  \includegraphics[width=0.99\linewidth]{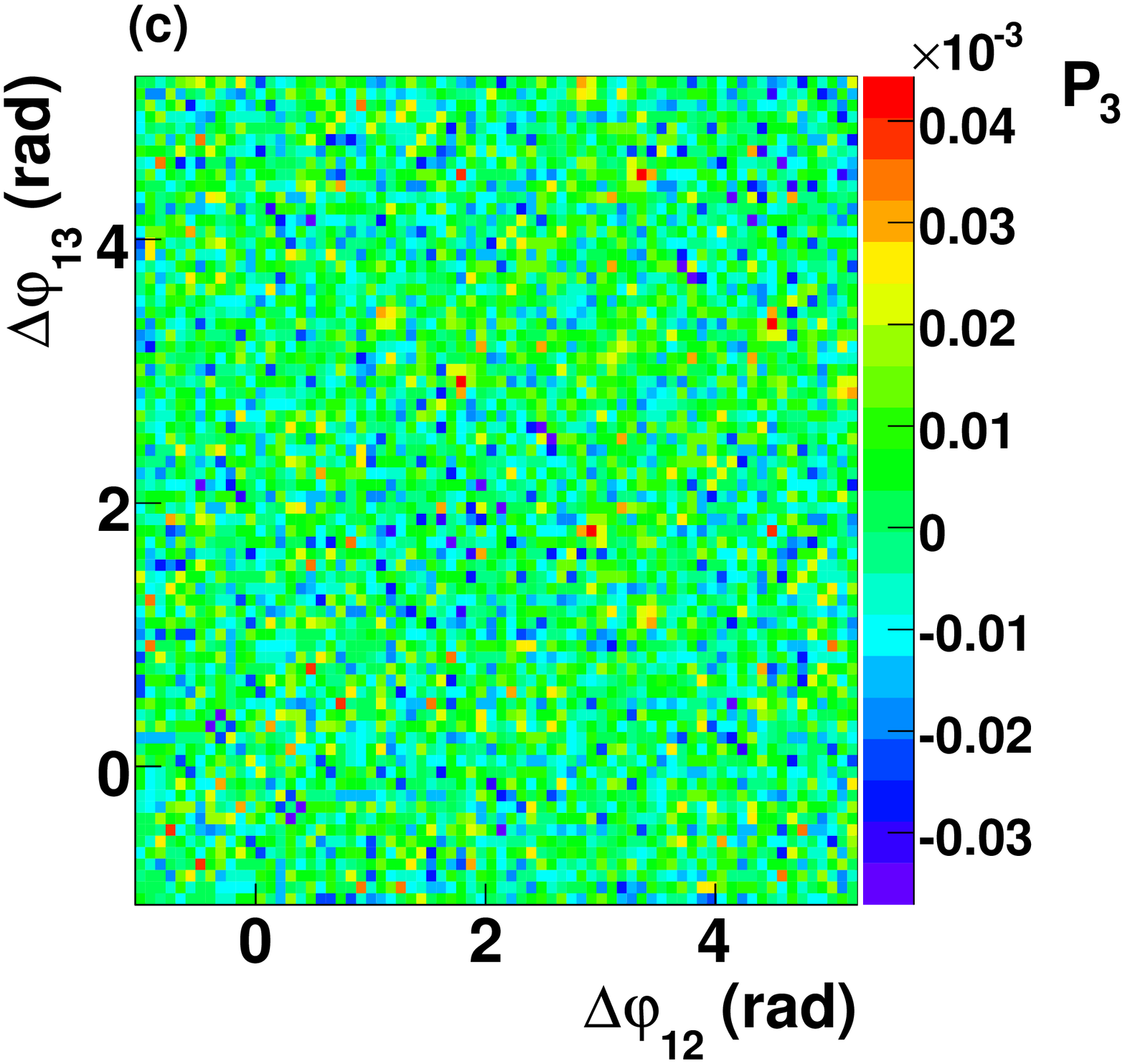}
  \end{center}
  \end{minipage}
}
\caption{Simulations of (a) the three-particle density, (b) three-cumulant, and (c) the 
probability cumulant for particle emission with elliptic anisotropy   $v_2 $ = 0.10.}
\label{Fig:FlowV2}
\end{figure}

Figure \ref{Fig:FlowV2} (a) and (b) respectively display the triplet density and the 3-cumulant 
obtained with the above flow random generator with  $v_2=0.1$, and  $v_4=0$.  The 3-cumulant, 
shown in Fig \ref{Fig:FlowV2} (b),  exhibits a finite  $v_2^2$ component owing to the fact 
that although the multiplicities  $N_i$  are generated with Poisson PDFs, the mean of these PDFs varies 
as a function of the event multiplicity thereby implying a correlation between the low and high $p_t$ particle multiplicities.
 This, in turn, implies  the coefficients  $f_{ijk}$
 are non-zero: the cumulant therefore contains a  non-vanishing $v_2^2$ component.  This component however 
vanishes in the probability cumulant shown in Fig \ref{Fig:FlowV2} (c) as expected from Eq. \ref{flowProbabilityCumulant}.

\begin{figure}[!htP]
\mbox{
  \begin{minipage}{0.33\linewidth} 
  \begin{center}
  \includegraphics[width=0.99\linewidth]{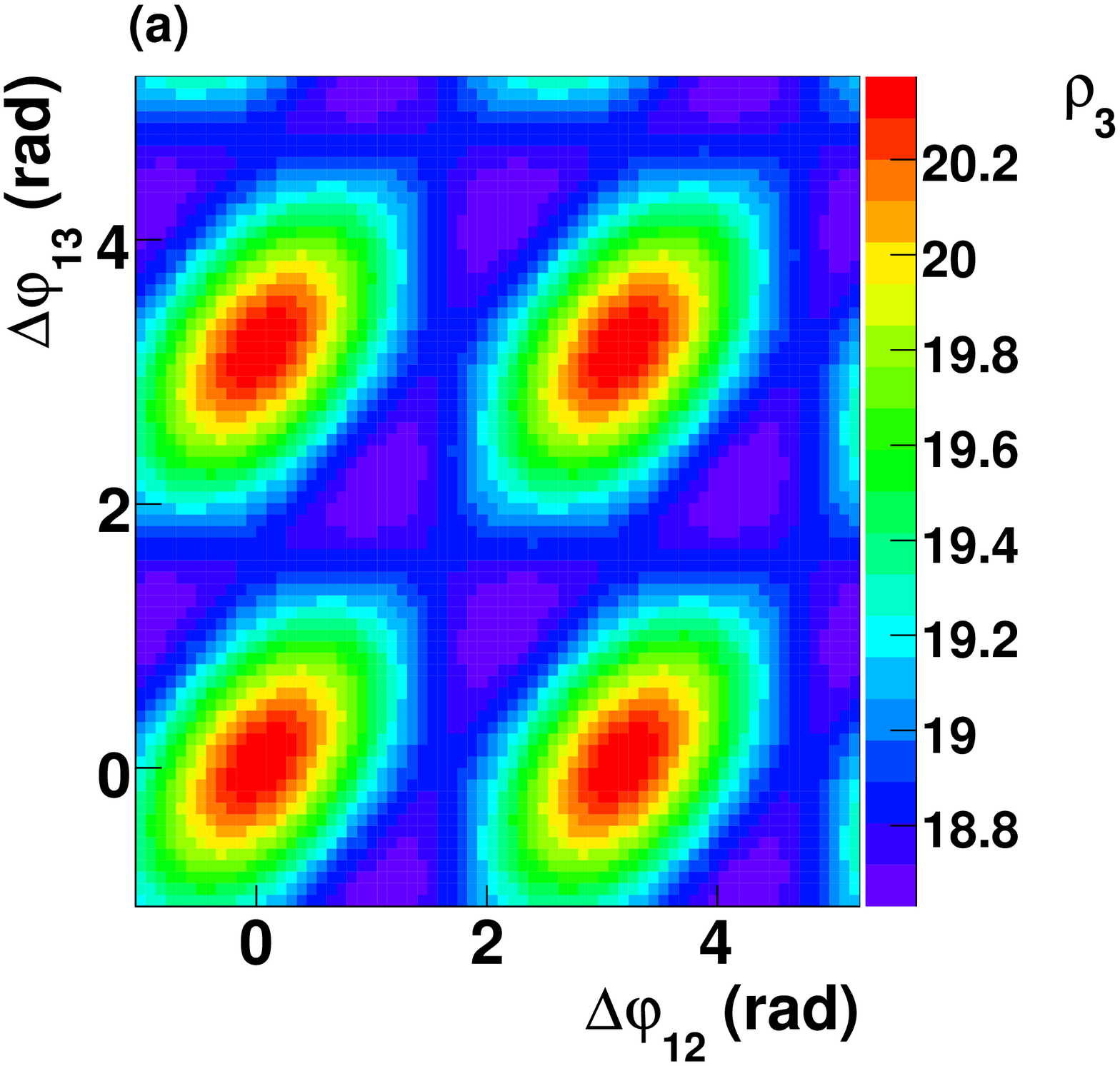}
  \end{center}
  \end{minipage}
  \begin{minipage}{0.33\linewidth} 
  \begin{center}
  \includegraphics[width=0.99\linewidth]{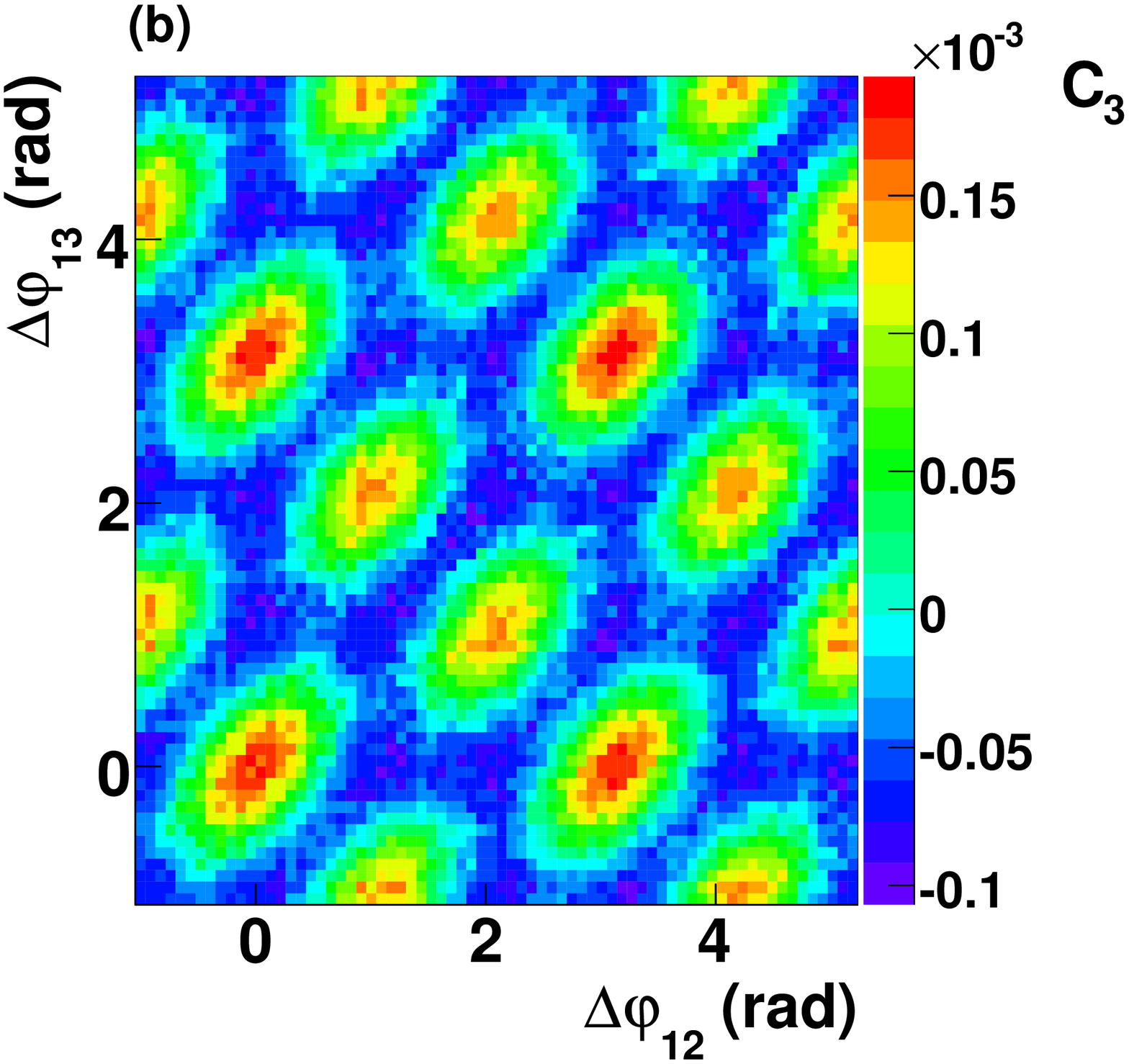}
  \end{center}
  \end{minipage}
  \begin{minipage}{0.33\linewidth}
  \begin{center}
  \includegraphics[width=0.99\linewidth]{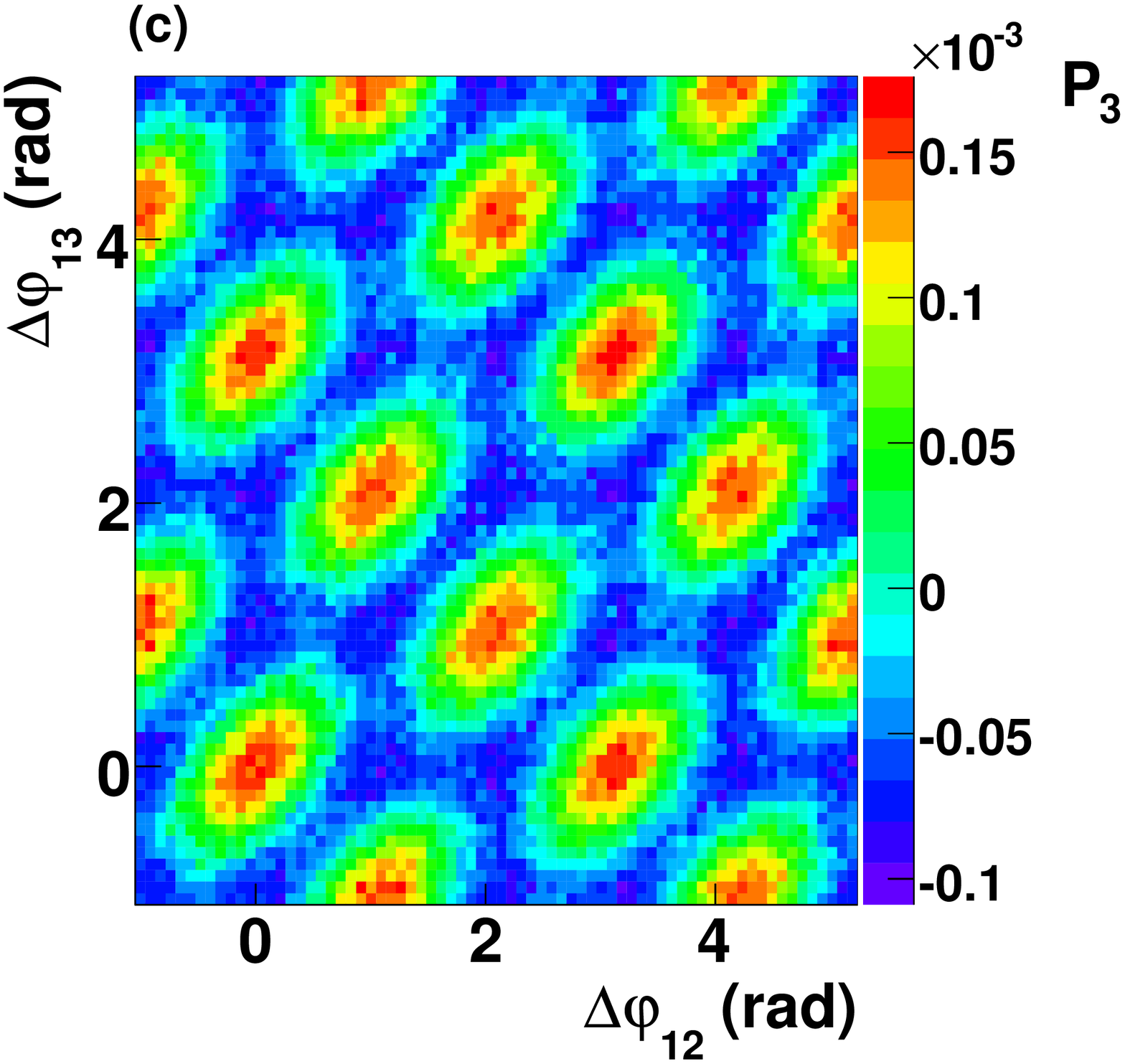}
  \end{center}
  \end{minipage}
}
\caption{Simulations of (a) the three-particle density, (b) three-cumulant, and (c) the 
probability cumulant for particle emission with elliptic anisotropy   $v_2= 0.10$, and  $v_4= 0.01$.}
\label{Fig:FlowV2V4} 
\end{figure}

Figure \ref{Fig:FlowV2V4} shows results of a simulation based on the same flow model as that used in 
Fig. \ref{Fig:FlowV2}, but with an added fourth harmonic component. Flow amplitudes are set to 
 $v_2=0.1 $ and  $v_4=0.01 $. While the 3-particle density (Fig \ref{Fig:FlowV2V4} (a)) is completely 
dominated by the 2$^{nd}$ harmonic, one finds the  3-cumulant suppresses this component 
significantly (Fig \ref{Fig:FlowV2V4}  (b)), and enables clear observation of the  $v_2^2v_4$ 
non-diagonal terms expected from Eq. \ref{probabilityCumulant}. The $v_2^2$ component is further suppressed in the 
probability cumulant, shown in Fig \ref{Fig:FlowV2V4}  (c), where only the  $v_2^2v_4$ component 
manifestly remains. As anticipated based on Eq. \ref{flowProbabilityCumulant}, the probability cumulant enables essentially 
full suppression of   $v_2^2 $ terms, and show irreducible flow components only.

Mono-jets and conical emission are next added to the simulated events. The jet axis (corresponding to the parton direction initiating 
the jet) is chosen randomly in the transverse plane. Particles are generated at random polar angles, 
relative to the jet direction using Gaussian PDFs. The width of the Gaussian is set to 0.15 radian for 
high  $p_t $ particles, and 0.25 for low  $p_t $ particles. Note that the conclusions of this study are 
essentially independent of the width of the jets. The associate multiplicity are generated jet-by-jet using 
Poisson PDFs.  The near side jet associated particle multiplicity (i.e. number of associates per jet) are 
set to 1 and 2 for the high- and low-$p_t $ particles respectively. No away side jet is included. Instead, one
introduces a conical emission as described in the following.
Jet parameters are selected to correspond approximately to associate jet yields measured by the CDF collaboration for jets of 10 to 
20 GeV \cite{CDF_JET}, as well as jet associated yields reported by RHIC experiments \cite{Horner06}. The number of jets is generated event-by-event with a Poisson PDF of mean 
0.01 jet per unit multiplicity: the average number of jets per event is thence of order 1-2.  Such large 
values are used to ease our study. Smaller, perhaps more realistic jet-yield values result in a weaker signal. This 
limits the statistical significance of the measurement, but does not constitute an intrinsic limitation of 
the cumulant method. 


In our simulations, the parton determining the direction of the jet is first generated, and 
is restricted for simplicity to the reaction transverse plane. The away-side  parton is emitted in the 
transverse plane as well, at angle of  $\pi $ radian relative to the jet parton. Conical emission is simulated 
by generating particles at a random polar angle relative to the away-side parton direction. The 
azimuthal direction of the particle, in the plane transverse to the parton direction, is randomly selected 
based on a uniform distribution. The polar angle is chosen to have a Gaussian distribution with an 
average value of  one (1) radian. The number of high-$p_t$ particles produced in this cone is arbitrarily 
set to zero, while the number of low-$p_t$ cone particles in a given parton-parton collision is selected 
randomly with a Poisson distribution of mean equal two, unless specified otherwise in the following.

\begin{figure}[!htP]
\mbox{
  \begin{minipage}{0.33\linewidth} 
  \begin{center}
  \includegraphics[width=1.\linewidth]{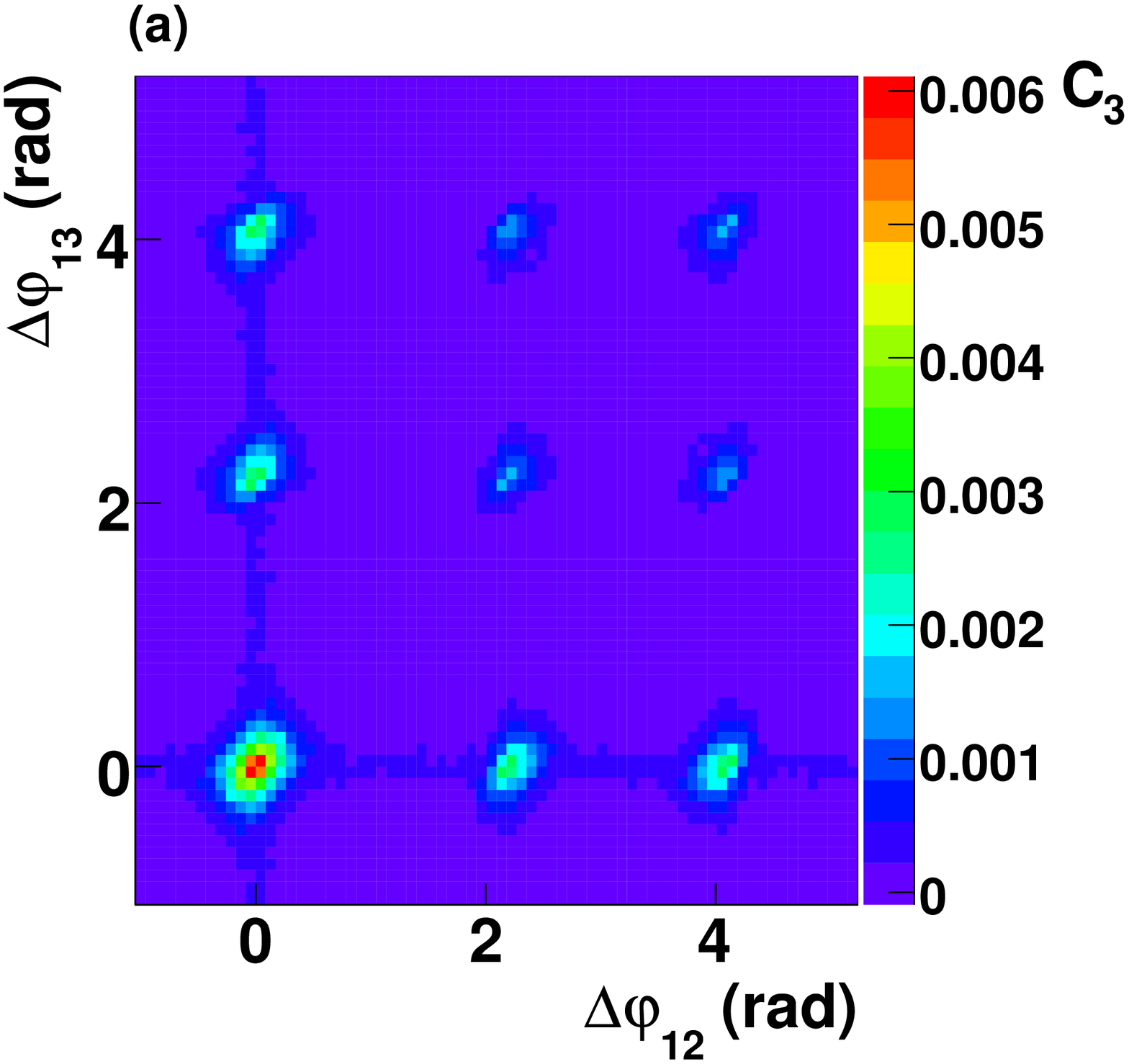}
  \end{center}
  \end{minipage}
  \begin{minipage}{0.33\linewidth} 
  \begin{center}
  \includegraphics[width=1.\linewidth]{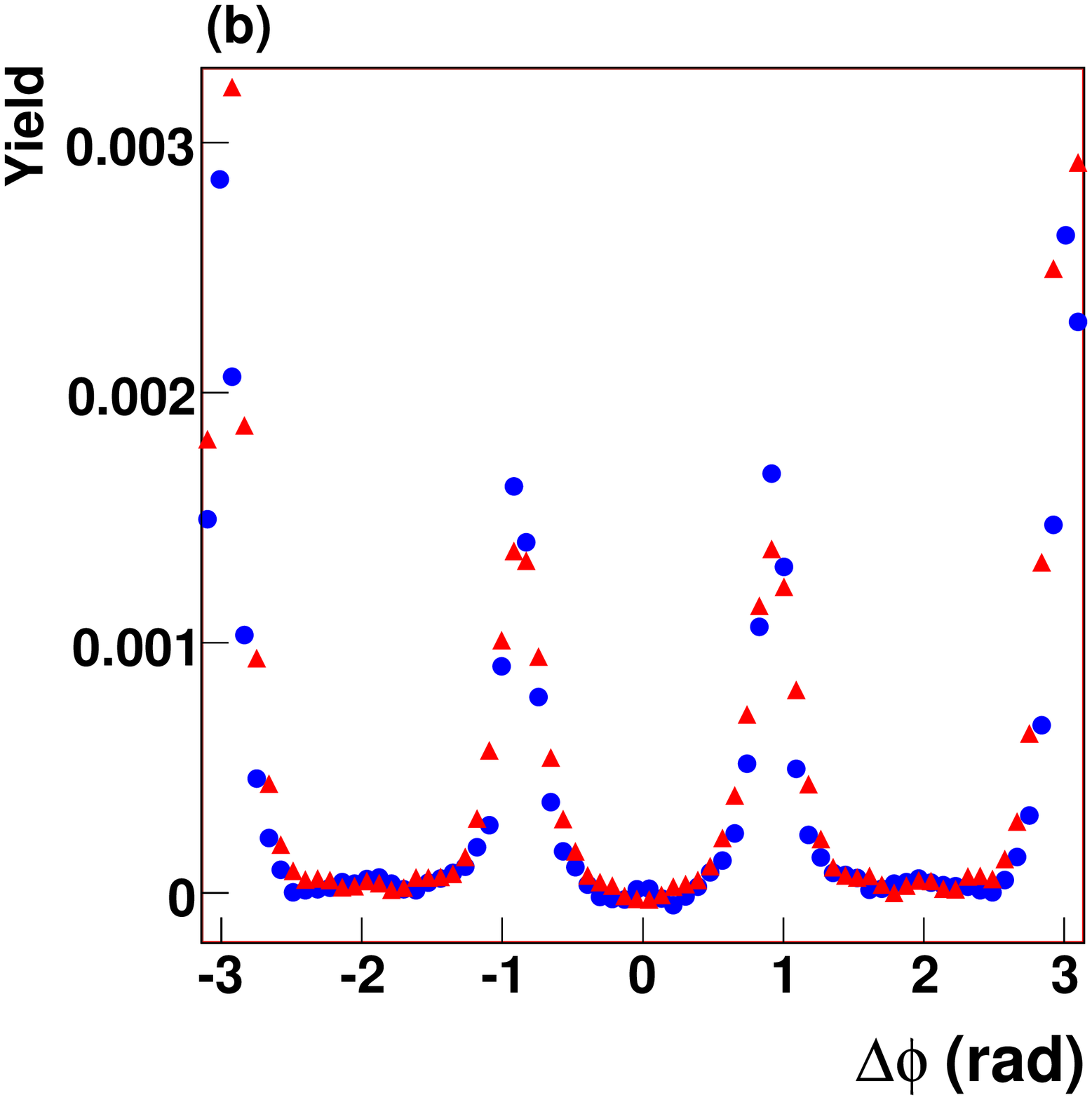}
  \end{center}
  \end{minipage}
  \begin{minipage}{0.33\linewidth} 
  \begin{center}
  \includegraphics[width=1.\linewidth]{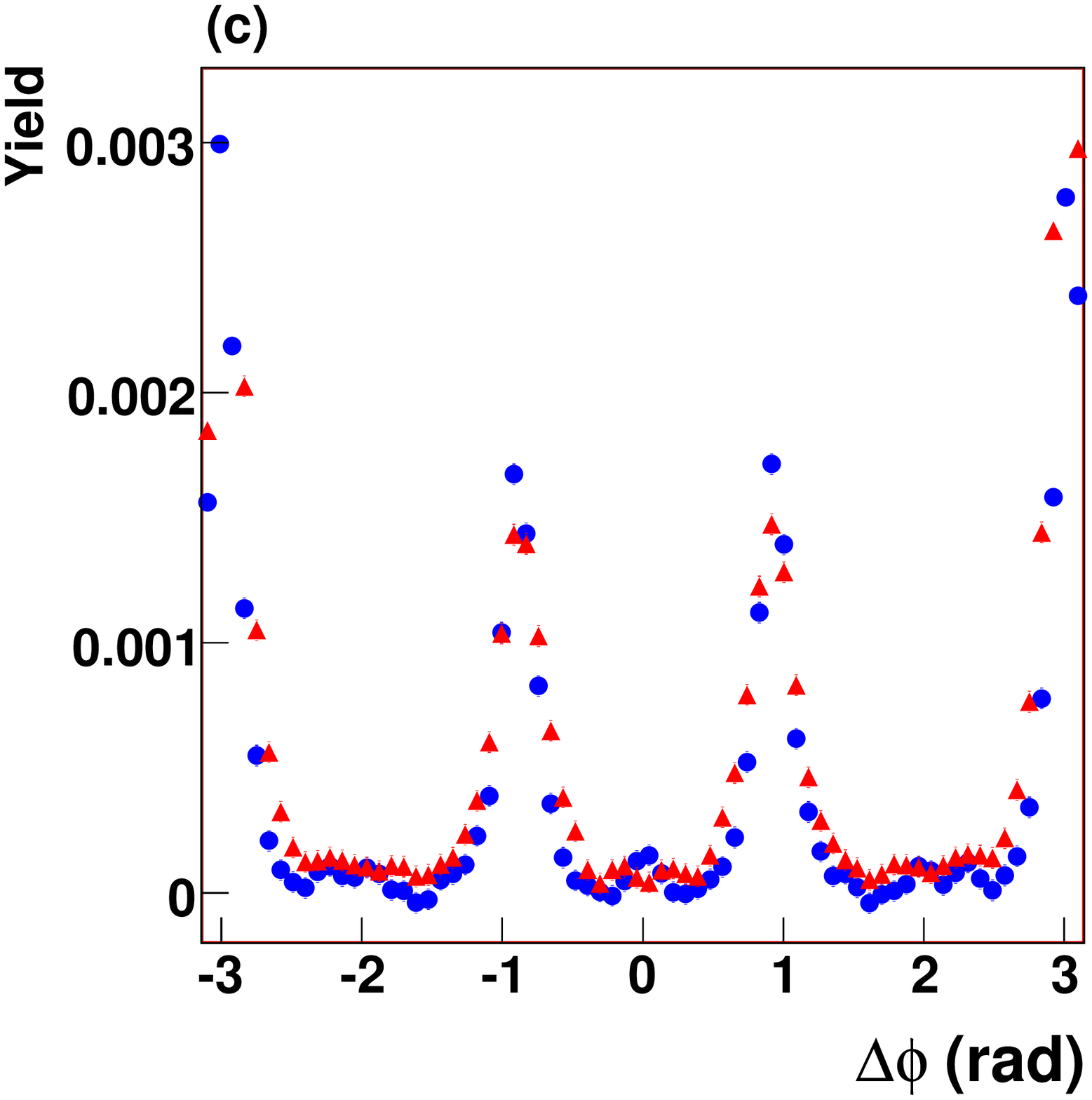}
  \end{center}
  \end{minipage}
}
\caption{(a) Simulation of the 3-cumulant obtained with  mono-jet, conical emission,  and 
background particles (without flow). Average jet yield to high-$p_t$ particle background yield set 
to 1.  (b)  Projections of the cumulant along the main (red) and alternate (blue) diagonals for jet/cone
parameters used in (a). Other particle generation parameters, and projections are discussed in the text. 
(c) Cumulant projections obtained for a high-$p_t$ background to jet-signal ratio of 4.
}
\label{Fig:Cone1} 
\end{figure}

Because the cone particles are emitted at an arbitrary polar angle, but detected in the transverse plane as 
a function of relative azimuthal angles only, conical emission results in four Jacobian peaks, as seen in 
Fig. \ref{Fig:Cone1} (a). In order to characterize the shape and strength of these peaks, we use 
projections of the cumulants along the diagonals  $\vphi_{12} + \vphi_{13}-2\pi $ and 
 $\vphi_{12}-\vphi_{13} $. Projections, shown in Fig. \ref{Fig:Cone1} (b) are limited to 
include narrow angular ranges about the diagonals. Specifically, the range  $|\vphi_{12}-\vphi_{13}|<10^o $
 is integrated to obtain the projection along the plot main diagonal (displayed in 
red), whereas the condition  $|(\vphi_{12}+\vphi_{13})-2\pi|<10^o $ is used for projections along 
 $\vphi_{12}-\vphi_{13} $ (shown in blue). 

\begin{figure}[!htP]
\mbox{
  \begin{minipage}{0.33\linewidth} 
  \begin{center}
  \includegraphics[width=1.\linewidth]{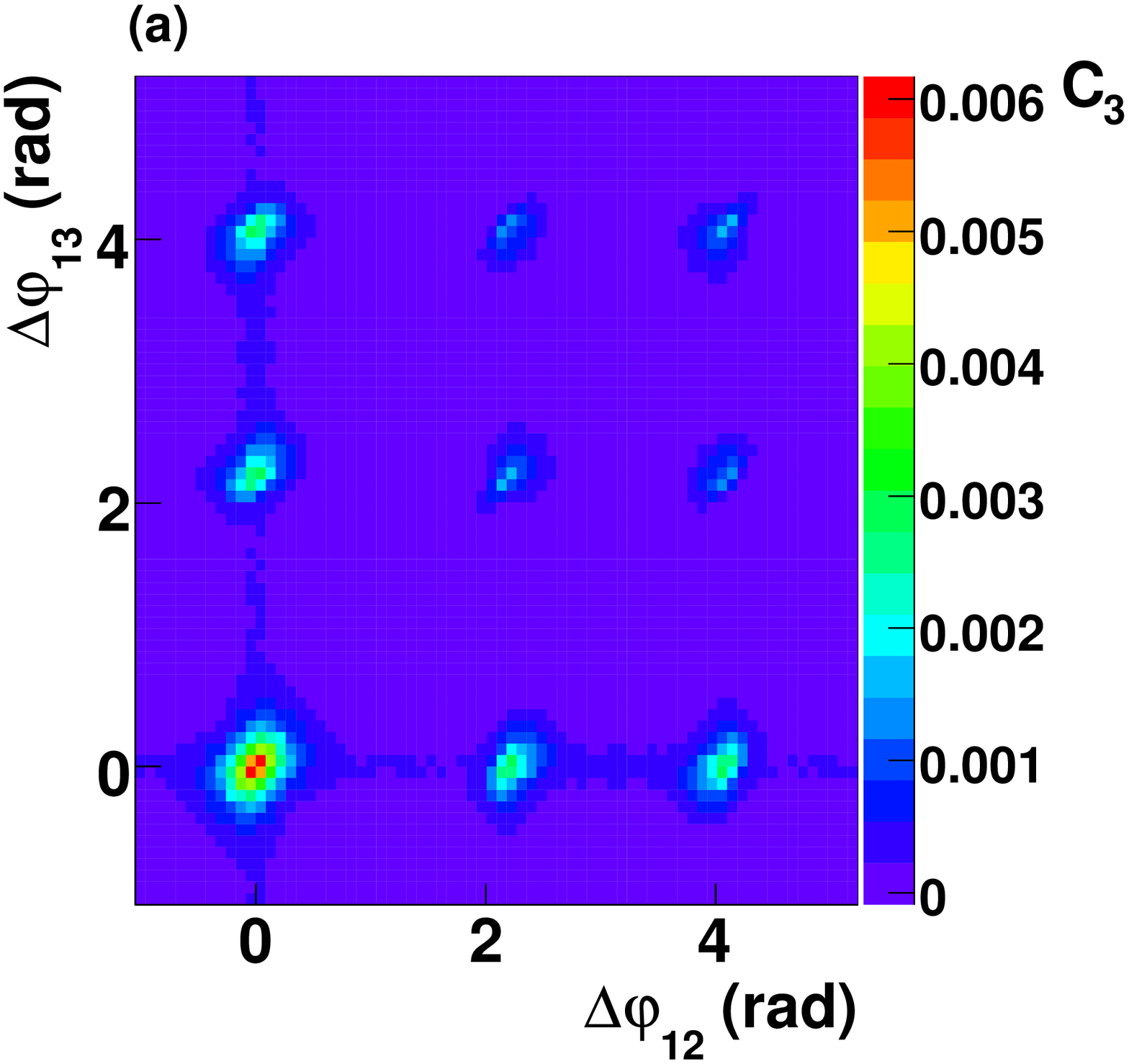}
  \end{center}
  \end{minipage}
  \begin{minipage}{0.33\linewidth} 
  \begin{center}
  \includegraphics[width=1.\linewidth]{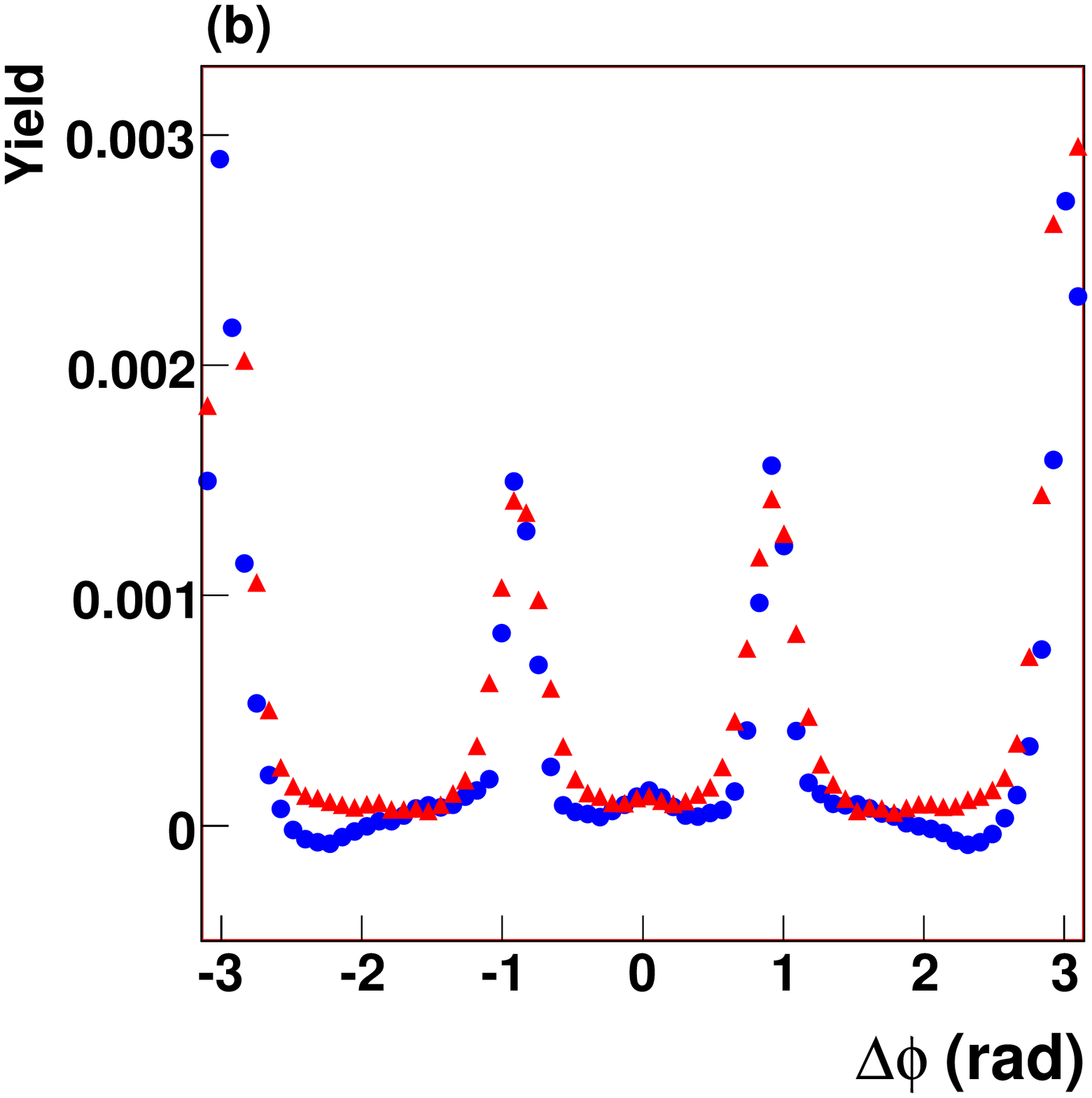}
  \end{center}
  \end{minipage}
  \begin{minipage}{0.33\linewidth} 
  \begin{center}
  \includegraphics[width=1.\linewidth]{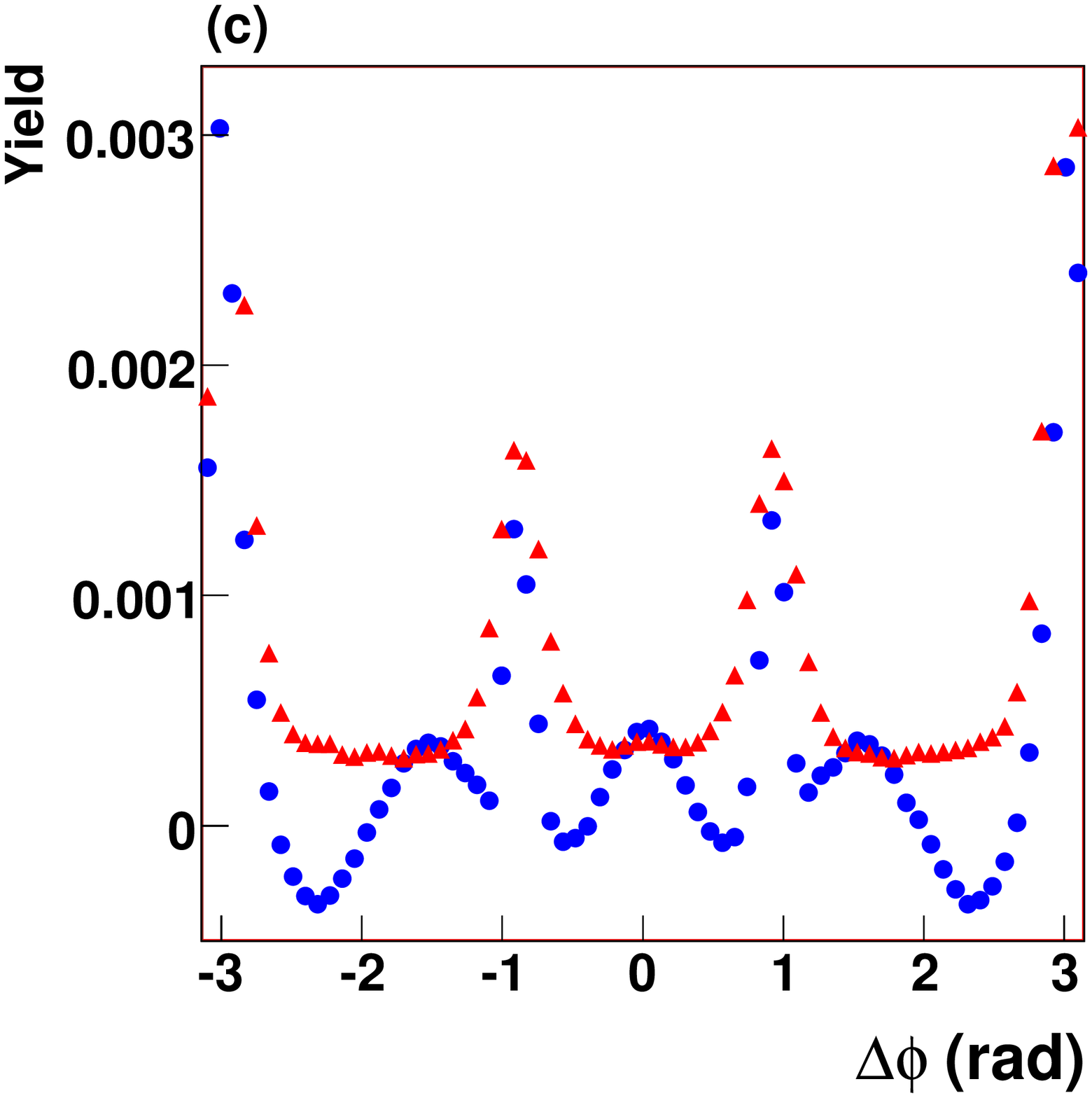}
  \end{center}
  \end{minipage}
}
\caption{Idem Fig. \ref{Fig:Cone1} in the presence of background flow.  (a-b) No background 
high-$p_t$ particle. Average of 100 low-$p_t$ background particle per event, with  flow  $v_2=0.1 $, 
and  $v_4=v_2^2 $.  (c)  Average of 100 low-$p_t$ background particle per event, with  flow  $v_2=0.2 $, 
and  $v_4=v_2^2 $.
}
\label{Fig:Cone3} 
\end{figure}

\begin{figure}[!htP]
\mbox{
  \begin{minipage}{0.33\linewidth} \begin{center}
  \includegraphics[width=1.\linewidth]{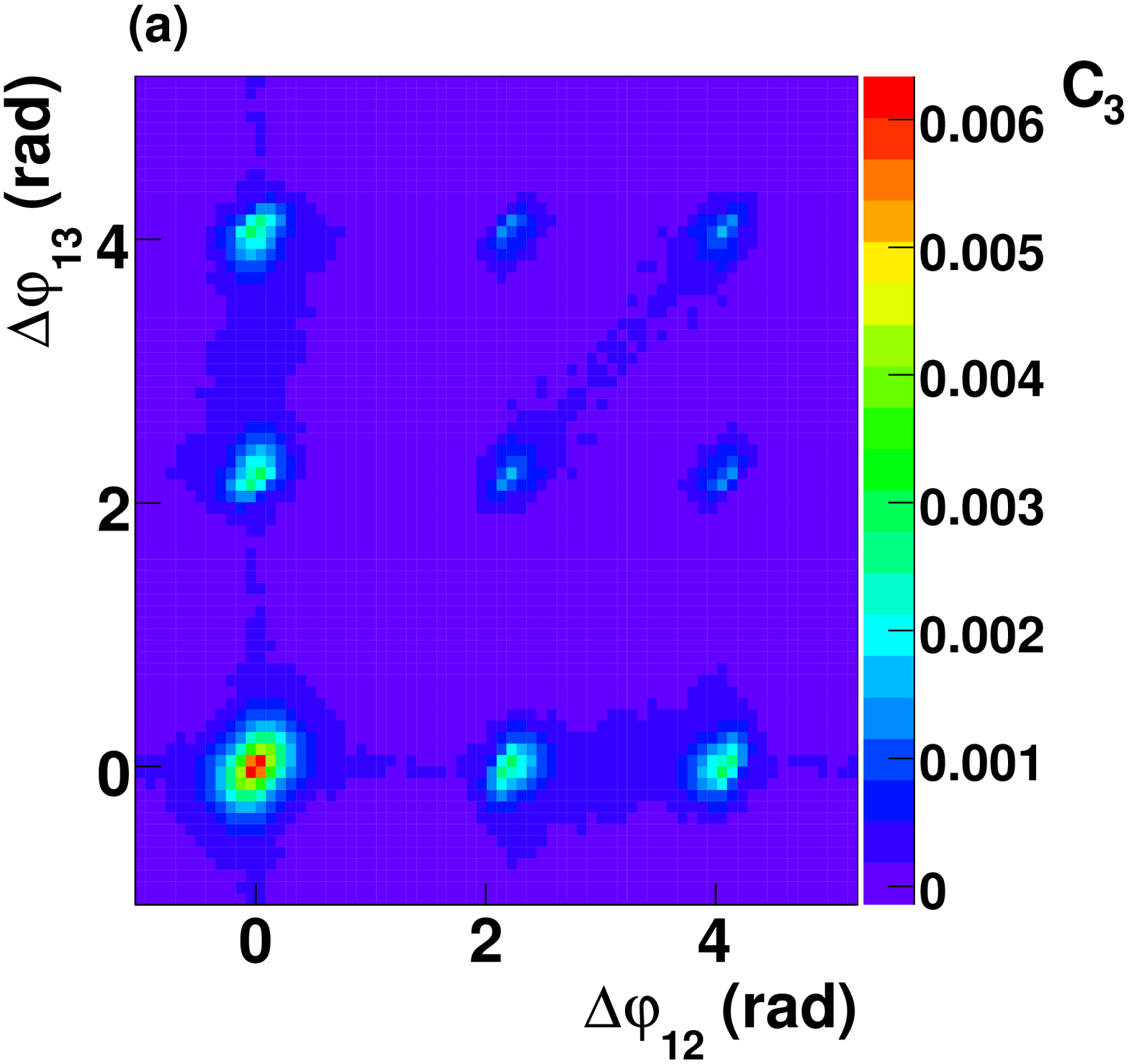}
  \end{center}
  \end{minipage}
  \begin{minipage}{0.33\linewidth} 
  \begin{center}
  \includegraphics[width=1.\linewidth]{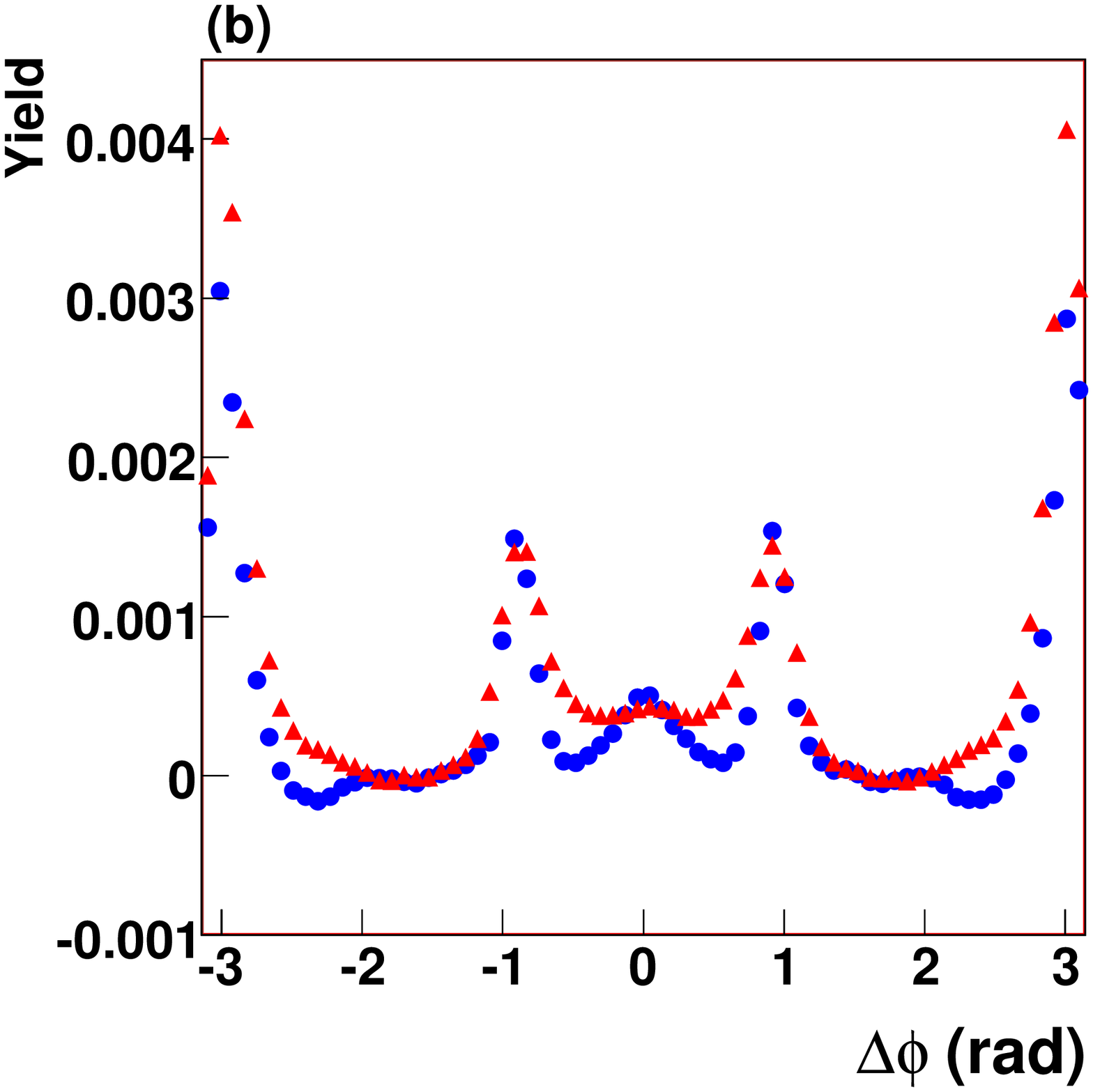}
  \end{center}
  \end{minipage}
    \begin{minipage}{0.33\linewidth} \begin{center}
  \includegraphics[width=1.\linewidth]{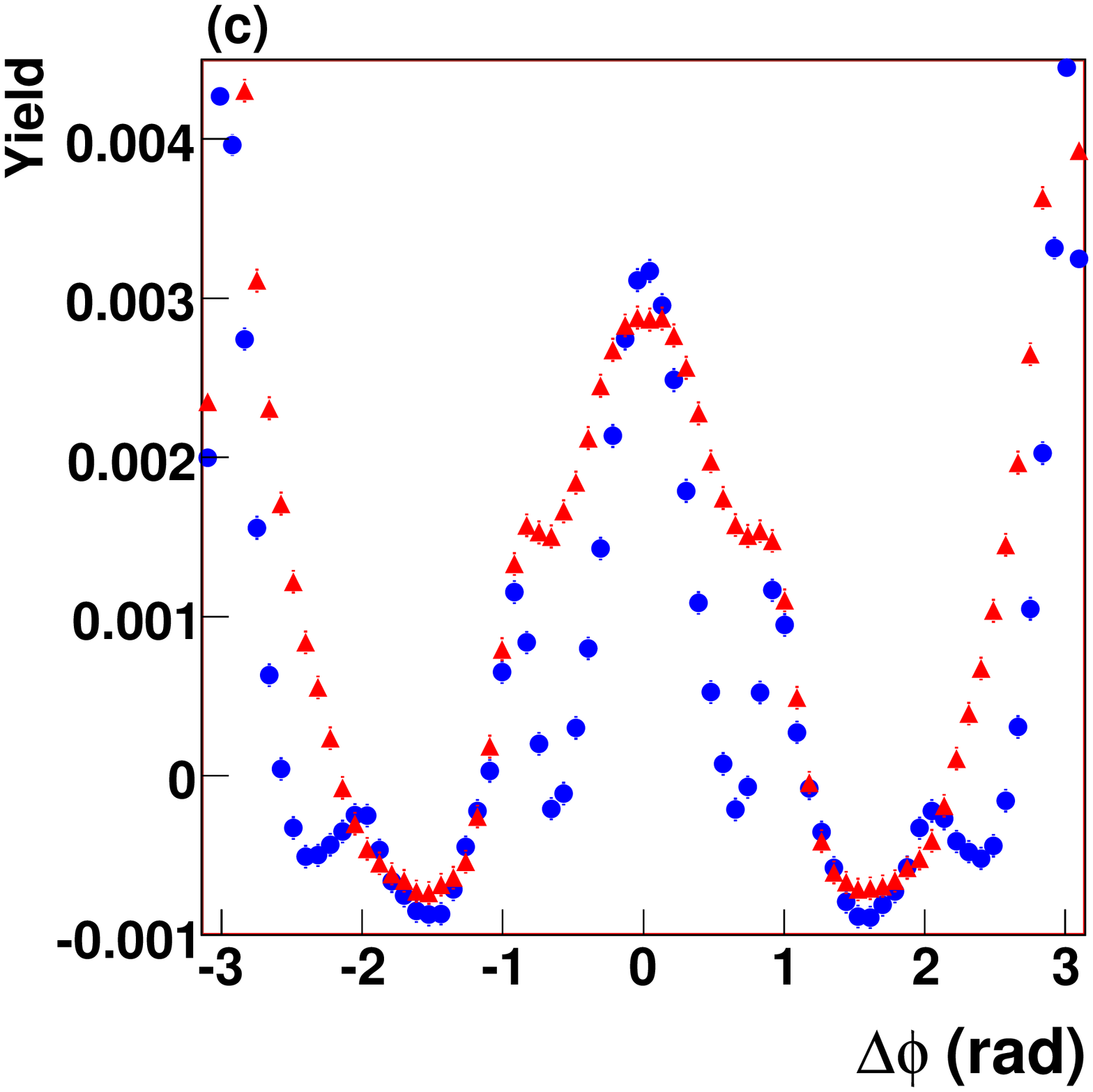}
  \end{center}
  \end{minipage}

}
\caption{(a-b) Idem Fig. \ref{Fig:Cone1}  for average jet yield to high-$p_t$ background equal one per event. 
Average of 100 low-$p_t$ background particle per event, with  flow  $v_2=0.1 $, and  $v_4=v_2^2 $. (c) Projections obtained for 
average jet yield to high-$p_t$ background equal eight (8) per event.
}
\label{Fig:Cone5} 
\end{figure}

\begin{figure}[!htP]
\mbox{
  \begin{minipage}{0.33\linewidth} 
  \begin{center}
  \includegraphics[width=1.\linewidth]{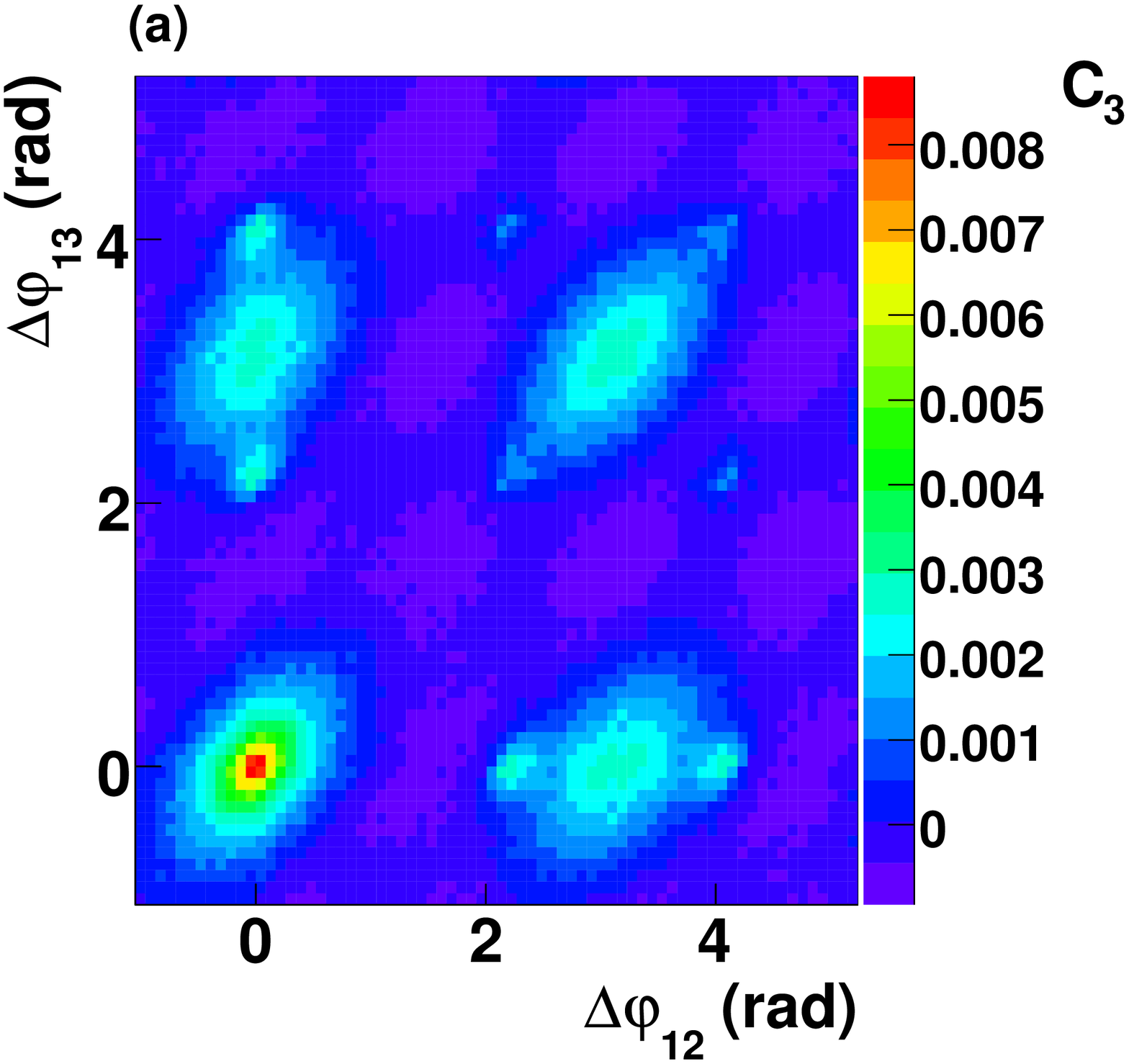}
  \end{center}
  \end{minipage}
  \begin{minipage}{0.33\linewidth} 
  \begin{center}
  \includegraphics[width=1.\linewidth]{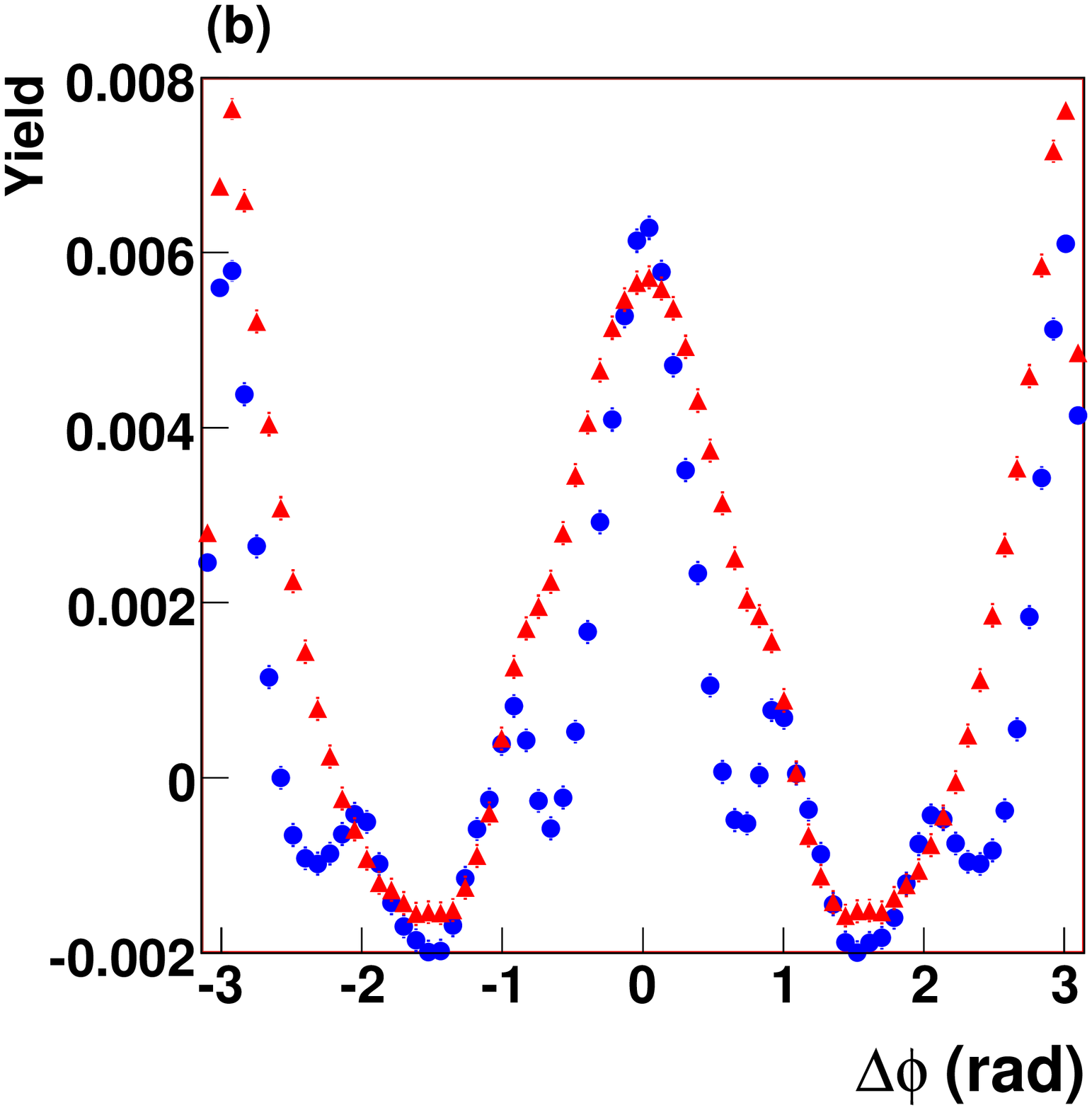}
  \end{center}
  \end{minipage}
}
\caption{Idem Fig. \ref{Fig:Cone1}  for average high-$p_t$ background to jet yield equal sixteen (16). 
Average of 100 low-$p_t$ background particle per event, with  flow  $v_2=0.1 $, and  $v_4=v_2^2 $.
}
\label{Fig:Cone8} 
\end{figure}

\begin{figure}[!htP]
\mbox{
  \begin{minipage}{0.5\linewidth} 
  \begin{center}
  \includegraphics[width=1.\linewidth]{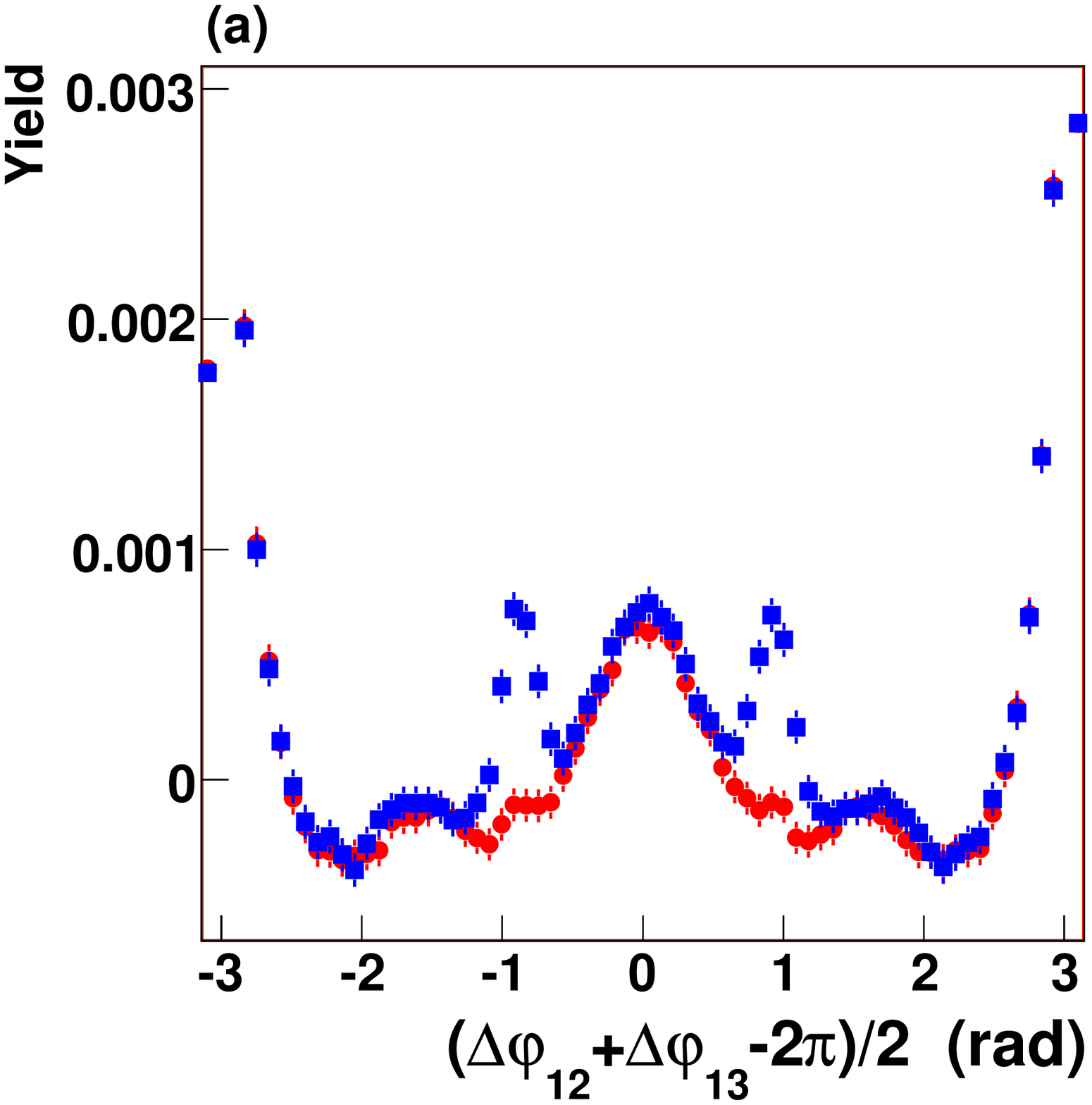}
  \end{center}
  \end{minipage}
  \begin{minipage}{0.5\linewidth} 
  \begin{center}
  \includegraphics[width=1.\linewidth]{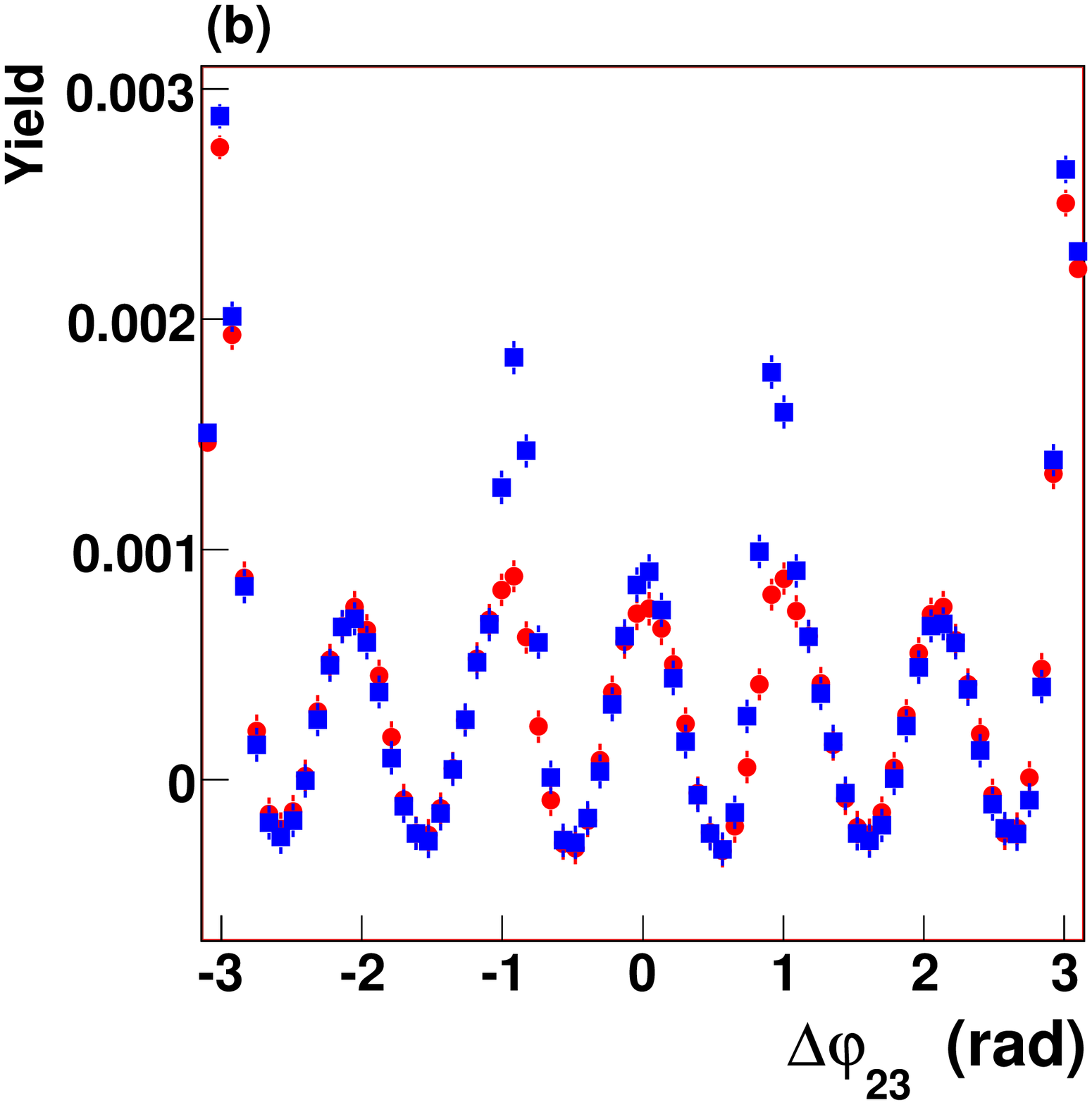}
  \end{center}
  \end{minipage}
}
\caption{(Color online) Comparison of the cone signal obtained with an average of two (blue squares) and one (red circles) low-$p_t$ particles involved
in conical emission. Panel (a) and (b) show the 3-cumulant projections along $(\Delta\vphi_12+\Delta\vphi_13)/2 $ and  $\Delta\vphi_{23}$ respectively.
 Simulations carried out with a jet-flow component, $v_2=0.05 $, and an average number of 
 100 low-$p_t$ background particles per event, with  flow  $v_2=0.1 $, and  $v_4=v_2^2 $. See text for details.
}
\label{Fig:Cone9} 
\end{figure}

At issue is whether or not one can extract a cone signal given the presence of background particles, and 
flow.  We thus vary the strength of the jet signal to background ratio by changing the fractional jet and 
background yields.  Figures \ref{Fig:Cone1} through \ref{Fig:Cone9} show cumulants and projections obtained with various model parameters. 
All plotted cumulants were obtained with simulations integrating two millions events.

Figures \ref{Fig:Cone1} (a-b) shows the cumulant and projections obtained for an average of 2.5 jets per event. The number of high and low  $p_t $ particles associated 
 with the jet, and cone are set respectively to 1 and 2. The uncorrelated background of high- and low-$p_t$ particle are set to 2.5 and 100 
particles per event respectively. The cone signal obtained with such parameters is obviously strong and clearly observable. 
Figure \ref{Fig:Cone1} (c) displays the cumulant and 
projections with a high-$p_t$ background raised by a factor of 4 resuling in a signal to noise ratio of 25\%.
The observed cone strength remains unchanged, as expected given the background is uncorrelated. 
We also verified that an increase of the width of the gaussian PDF used for jet, and cone particle emission results in wider 
peak structures with reduced amplitude but no actual change in the integrated cone signal strenght. In this context, we conclude the observability of conical emission is only limited
by the statistical accuracy of the measurement relative to the actual strength of the signal. For instantce, a reduction by a factor of five of the number of high-$p_t$ 
associated particles results in cone-signal five times small in amplitude, but the shape of the cumulant remains unchanged and the observability of the signal is thus only limited
by the size of the event sample, and the number of particles associated with the trigger and cone.

We next consider the observability of conical emission in the presence of anisotropic flow. Figure \ref{Fig:Cone3} shows the cumulant and projections 
in the presence of a flow background. The jet and conical signals are identical to those used in Fig. \ref{Fig:Cone1} (a-b). An average background of 100 low-$p_t$ particles is included with 
 $v_2=0.1 $, and  $v_4=v_2^2 $. No high-$p_t$ background is included. The cone signal shape and strength remains unaffected, although the projections exhibit
a background structure somewhat more complex than that observed in Fig. \ref{Fig:Cone1}. Fig. \ref{Fig:Cone3} (c) displays the  projections 
obtained when the elliptic flow is raised to  $v_2=0.2 $. While the signal is clearly visible, one observes 
the emergences of cosine structures along the  $\Delta\vphi_{23} $ projection. These stem from the non-Poisson nature of the background multiplicity fluctuations used 
in the simulations.

Flow-like or cosine structures also appear when high-$p_t$ background particles are added as shown in Figure \ref{Fig:Cone5} where the average high-$p_t$ particle background is set to 2.5/event.
The observability of the cone depends on its amplitude relative to that of the background, as well as on the magnitude of  $v_2 $ and  $v_4 $. We note that the signal remains
visible in the  $\Delta\vphi_{23} $ projection, with the jet and cone parameters used in Figure \ref{Fig:Cone1}, even when the high-$p_t$ background is raised by a factor of eight as shown in Fig. 
\ref{Fig:Cone5}. For substantially larger background, the signal however becomes more difficult to extract as illustrated in
Figure \ref{Fig:Cone8} where a background to signal ratio of sixteen is used for high-$p_t$ particles.
Clearly,  the observability of a cone signal in the presence of flow depends on the strength of the signal relative to that of non-Poisson components and values of $v_2$.

\begin{figure}[!htP]
\mbox{
  \begin{minipage}{0.5\linewidth} 
  \begin{center}
  \includegraphics[width=1.\linewidth]{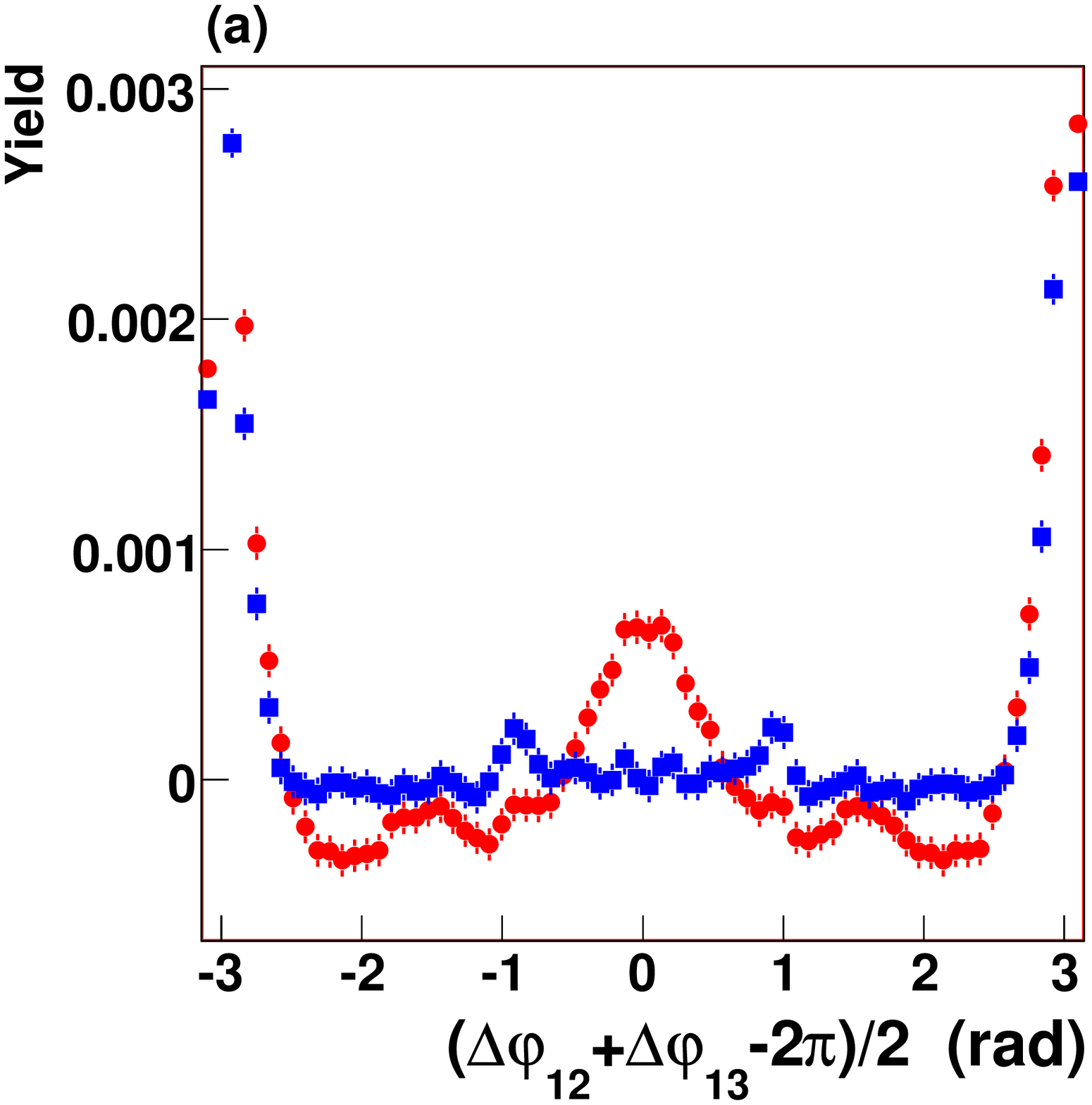}
  \end{center}
  \end{minipage}
  \begin{minipage}{0.5\linewidth} 
  \begin{center}
  \includegraphics[width=1.\linewidth]{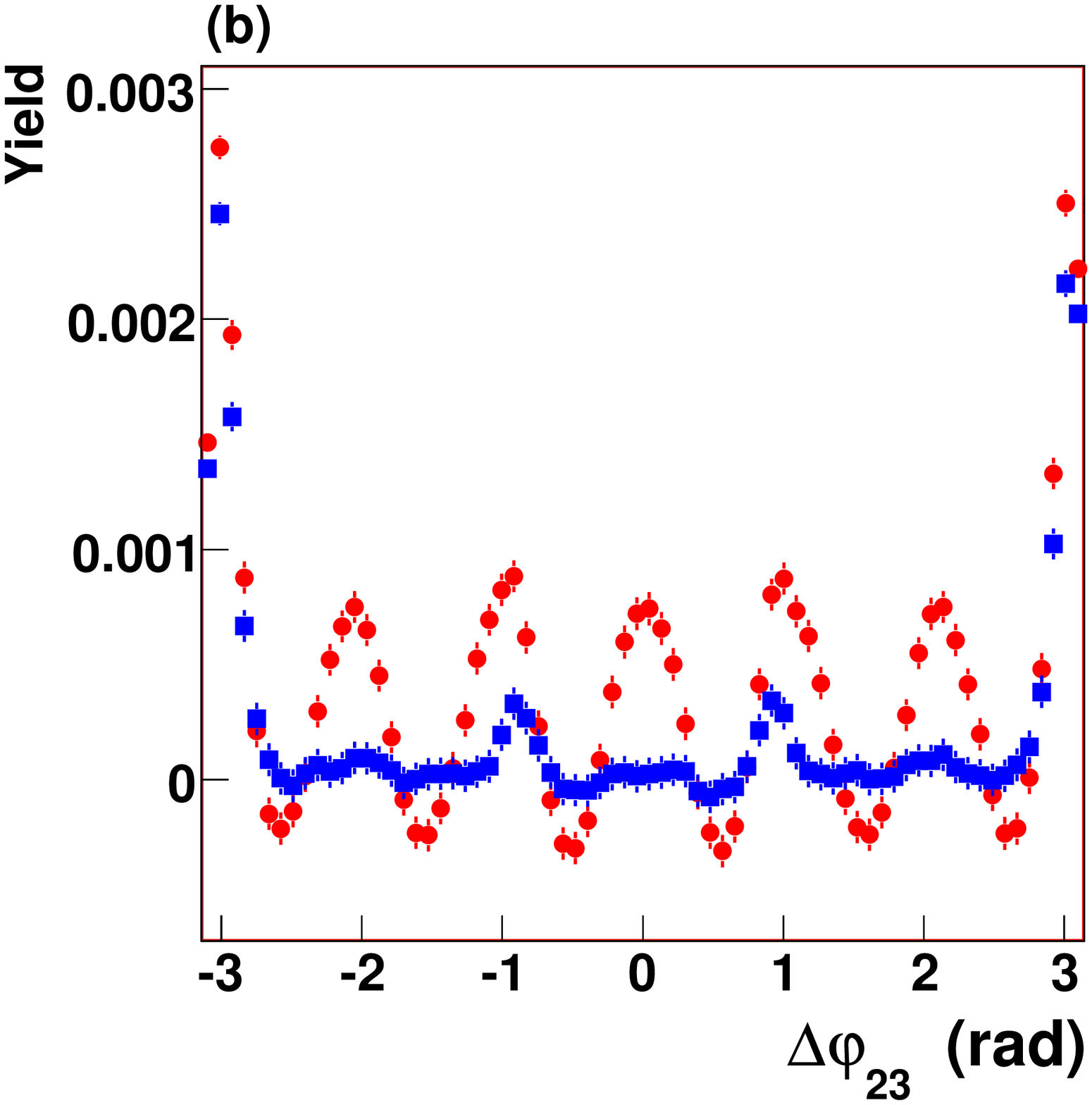}
  \end{center}
  \end{minipage}
}
\caption{(Color online) Comparison of the cone signal obtained with an average of one  low-$p_t$ particles involved
in conical emission and an average of 100 low-$p_t$ background particles per event with $v_2=0.05 $ (red circles)  and $v_2=0.1 $(blue squares). 
Panel (a) and (b) show the 3-cumulant projections along $(\Delta\vphi_12+\Delta\vphi_13)/2 $ and  $\Delta\vphi_{23}$ respectively.
 Simulations are carried out with a jet-flow component, $v_2=0.05 $. The background has $v_4=v_2^2$.
}
\label{Fig:Cone10} 
\end{figure}

The model independent extraction of a conical signal becomes increasingly difficult as the signal strength is reduced and the elliptic flow magnitude increased. The 
measurement becomes particularly challenging when the jet amplitude is modulated by differential attenuation relative to the reaction plane,
 i.e. in the presence of jet-flow. Figures \ref{Fig:Cone9} (a) and (b) display projections along  $\Delta\vphi_12+\Delta\vphi_13 $ and $\Delta\vphi_23$ obtained
 with  jet  $v_2=0.05 $, (as defined in Sec. \ref{FlowJetCorrelations}).
The blue squares and red circles show, respectively, the  signals obtained with an average of two  and one low-$p_t$ particles involved
in conical emission. 
Simulations are carried out with an average number of 2.5 jets/cones per event. The average yield of high-$p_t$ low-$p_t$ background particle are set to 8 and 100 
per event respectively. Background particles are produced with  flow  $v_2=0.1 $, and  $v_4=v_2^2 $. In our simulation, the number of cone associated particles is
determined  event-by-event with a Poisson random number generator. The conical signal is  detectable, in the context of this 3-cumulant analysis, 
only when two, or more,  particles are generated per event. For an average cone yield of two associated particles,  60\% of 
the cone events contain two or more cone particles. By contrast, only 26\% of the events contain two cone particle when the average (cone) yield is 
one. The signal obtained in the former case is clearly visible, while if becomes weaker in the latter. The detectability of the signal thus 
greatly depends of the actual yield of conical emission. We note that two-particle correlation analyses report an away side yield of two particle per trigger
after flow background subtraction based on a ZYAM hypothesis \cite{ZYAM}. If  this away side is due to conical emission, this implies the low-$p_t$ cone
particle average multiplicity is large and should therefore be detectable in the context of a 3-cumulant analysis. 
We note however that the cumulant measurements reported by the STAR collaboration at recent conferences
do not exhibit as strong  $cos(2\Delta\vphi) $ components as those illustrated in  the  $\Delta\vphi_1+\Delta\vphi_2 $ projections shown in Fig. \ref{Fig:Cone9}
\cite{Pruneau06a}. Note additionally that the cone signal
remains visible in the  $\Delta\vphi_{23} $ projection even  for small signal to background ratios. It is however obvious that strong background flow and jet-flow
may hinder the observation of conical emission. For instance, Figure \ref{Fig:Cone10} shows a comparison of the 3-cumulant projections obtained for conical emission
in the presence of jet-flow cross terms with $v_2$ values of 0.05 (blue) and 0.1 (red) an associate cone yield of one. One finds the cone single is difficult 
to distinguish for $v_2=0.1$ but straightforward to observe for $v_2=0.05$.

In summary, e find in the context of our multicomponent model that the exact sensitivity of the measurement is  a "complicated" function 
of the flow and background parameters of the model, and cannot readily be expressed in analytical form or even as a table. We further note that 
the shape and strength of actual correlations is influenced by momentum, and quantum number conservation effects not explicitly addressed in this work. 

\section{Summary and Conclusions}
\label{Sect:Conclusion}

Conical emission has been suggested in the recent literature to explain the away-side structures observed in two particle azimuthal correlations
for $Au+Au$ collisions at $\sqrt{s_{NN}}=200$ GeV. If realized, conical emission should also lead to three particle correlations. 
We introduced in Ref. \cite{Pruneau06} a method based on cumulants to carry a search for such a signal. In this work, we presented 
a discussion of the properties of the three particle cumulant (3-cumulant) calculated in terms of two relative azimuthal angles,
as defined in \cite{Pruneau06}, for searches of conical emission in high-energy 
$A+A$ collisions. We showed that in the presence of azimuthal anisotry (flow), the 3- cumulant reduces to  non-diagonal terms 
dominated by components of order $v_2v_2v_4$ for Poissonian particle production. Given particle production is in general non-Poissonian,
one however expects the presence of second order terms, dominated by components in $v_2^2$ in the cumulant. The strength of those terms relative
to the irreducible  $v_2v_2v_4$ components depends on the magnitude of $v_2$ and the global strength of particle correlations, or departure 
from Poisson statistics. The strength of particle correlation is known to vary inversely with the collision multiplicity in $A+A$ collisions
due to increasing two-particle correlation dilution with increasing number of collision participants. We thus expect the mix of $v_2v_2v_4$
and $v_2^2$ should be a function of collision centrality. 

We introduced a three-particle probability cumulant, and showed it is devoid 
of $v_2^2$ non-Poissonian components in the presence of flow only thereby enabling, in principle, the determination of $v_2v_2v_4$ amplitudes. We discussed
the shape and strength of the 3-cumulant based on simple particle production models including di-jets, mono-jet plus conical emission, and 
jet$\times$flow correlations. We showed that jet$\times$flow cross-terms should be more complex than those assumed by some
ongoing analyses, and cannot, in particular, be derived from a product of two-particle flow and jet-like terms. 
We used the models to discuss the sensitivity of the 3-cumulant for searches of conical emission
in $A+A$ collisions. We showed that the cumulant enables excellent sensitivity to conical emission for modest values of flow $v_2$, but
becomes increasingly challenging for larger values of $v_2$. We further showed that conical emission can be detected even in the presence
of jet$\times$flow correlations for small values of $v_2$ jet flow. It is however obvious the signal may be masked by large values of jet and background 
$v_2$ and large non-Poissonian particle production. Precise values of sensitivity, in terms of signal strength to background ratios, cannot be expressed in a model
independent way, and are found to depend on the relative amplitude of background $v_2$ and $v_4$ jet-flow correlations, as well as the global
strength of particle correlations, i.e. departure from Poisson statistics.   

This work neglects effects associated with
momentum conservation, quantum number conservation, and particle correlations present in $p+p$ collisions. Momentum conservation implies
in particular that a multicomponent description of particle production is not strictly valid, and while it is useful to estimate the effects
of various particle production mechanisms, its use to search for conical emission signals is model dependent and therefore unreliable.

{\bf Acknowledgements}

The author thanks S. Voloshin, M. Sharma, and A. Poskanzer for their critical reading of the manuscript, valuable comments and discussions. This work was supported in part by DOE Grant No. DE-FG02-92ER40713.


\section*{References} 
\begin{thebibliography}{10} 


\bibitem{Stoecker76} J. Hoffmann, H. Stoecker, U. Heinz, W. Scheid, and W. Greiner, Phys. Rev. Lett. {\bf 36}, 88 (1976).

\bibitem{Stoecker05}  H. Stoecker, Nucl. Phys. {\bf A750}, 121 (2005).

\bibitem{JRuppert05} J. Ruppert, J. Phys. Conf. Ser. {\bf 27}, 217 (2005).

\bibitem{SolanaShuryak05}  J. Casalderrey-Solana, E. V. Shuryak, and D. Teany, J. Phys. Conf. Ser. {\bf 27}, 22 (2005).

\bibitem{RupperMuller05}  J. Ruppert, B. MŸller, Physics Letters  {\bf B618}, 123 (2005).

\bibitem{RenkRuppert06} T. Renk and J. Ruppert, Phys. Rev. C {\bf 73}, 011901 (2006).


\bibitem{Heinz06}  A. K. Chaudhuri, U. W. Heinz, Phys. Rev. Lett. {\bf 97}, 062301 (2006).

\bibitem{Stocker07}  H. Stocker, {\em et al.}  nucl-th/0703054, 2007.

\bibitem{RenkRupper07}  T. Renk, J. Ruppert, hep-ph/0702102.

\bibitem{MullerQM08} B. Muller, J. of Physics G: Nucl. Part. Phys. (QM 08 Proceedings) in press.

\bibitem{BBetz08} B. Betz, J. of Physics G: Nucl. Part. Phys. (QM 08 Proceedings) in press.



\bibitem{Mukherjee07} S. Mukherjee, M. Mustafa, and F. Ray, Phys. Rev. D {\bf 75}, 094015 (2007).


\bibitem{Gubser08} S. Gubser, S. Pufu, and A. Yarom, Phys. Rev. Lett. {\bf 100} 012301. (2008).


\bibitem{Dremin79}   I.M. Dremin, JETP Lett. {\bf 30} 140 (1979).

\bibitem{Dremin81}   I.M. Dremin, Sov. J. Nucl. Phys.{\bf 33} 726 (1981).

\bibitem{Dremin06}   I.M. Dremin, Nucl. Phys. A {\bf 767}, 233 (2006).

\bibitem{KochMajumber05} V. Koch, A. Majumber, X.-N. Wang, Phys. Rev. Lett. {\bf 96}, 172302 (2005). 

\bibitem{MajumderMuller06} A. Majumder, B. Muller, S. A. Bass (2006), hep-ph/0611135.


\bibitem{Vitev05}  I. Vitev, Phys. Lett. B {\bf 630}, 78 (2005).

\bibitem{PolosaSalgado05}   A.D. Polosa and C.A. Salgado, hep-ph/0607295. 


\bibitem{Hwa05}  C. B. Chiu, R. C. Hwa, Phys. Rev. C{\bf 74}, 064909 (2006).



\bibitem{Armesto04}   N. Armesto, {\em et al.}, Phys. Rev. Lett. {\bf 93}, 242301 (2004), hep-ph/0405301.


\bibitem{Salgado05a}  N. Armesto, C.A. Salgado, U. A. Wiedemann, Phys. Rev. C {\bf 72}, 064910 (2005).

\bibitem{RenkRuppert07_PLB646} T. Renk and J. Ruppert, Phys. Lett. {\bf  B646}, 19 (2007).

\bibitem{RenkRuppertPRC2007} T. Renk and J. Ruppert, Phys. Rev. {\bf  C76}, 014908 (2007).

\bibitem{Voloshin05} S.A. Voloshin, Phys. Lett. B {\bf 632}, 490 (2006) arXiv:nucl-th/0312065.

\bibitem{Hwa05_PRC74}  C. B. Chiu, R. C. Hwa, Phys. Rev. C{\bf 72}, 034903 (2005), nucl-th/0505014.


\bibitem{Holzmann05}  W. G. Holzmann, N. N. Ajitanand, J. M. Alexander, P. Chung, M. Issah, R. A. Lacey, A. Taranenko and A. Shevel, J. Phys. Conf. Ser. {\bf 27}, 80 (2005).



\bibitem{Ajitanand06}  Chun Zhang {\em et al.}  (PHENIX Collaboration), J. of Physics G: Nucl. Part. Phys. {\bf34}, S671 (2007).


\bibitem{Ulery06_774}  J. Ulery, {\em et al.}  (STAR Collaboration), Nucl. Phys. A {\bf 774}, 581 (2006).

\bibitem{Ulery06a}  J. Ulery and F. Wang, nucl-ex/0609016.

\bibitem{Ulery06b}  J. Ulery, {\em et al.}  (STAR Collaboration), nucl-ex/07040224.

\bibitem{Ulery06c}  J. Ulery, Ph.D. Thesis, Purdue University, (2007).

\bibitem{Ajitanand06_774}N.N. Ajitanand (PHENIX Collaboration), Nucl. Phys. {\bf A774} 585 (2006).


\bibitem{Ulery07}  J. Ulery, {\em et al.}  (STAR Collaboration), Nucl. Phys. A {\bf 783}, 511 (2007).

\bibitem{CeresQM06}  Stefan Kniege and Mateusz Ploskon, {\em et al.}  (CERES  Collaboration), J. of Physics G: Nucl. Part. Phys. {\bf 34}, S697 (2007).

\bibitem{Kniege07} S. Kniege, M. Ploskon (CERES Collaboration), nucl-ex/0703008.

\bibitem{Ajitanand07} N. N. Ajitanand, Nucl. Phys. {\bf A783} 519 (2007).


\bibitem{Borghini06}  N. Borghini, Phys.Rev. C75 (2007) 021904.

\bibitem{Lisa08} Zbigniew Chajecki, and Mike Lisa, arXiv:0807.3569.

\bibitem{Pruneau06} C. A. Pruneau, Phys. Rev. {\bf C74} 064910 (2006).

\bibitem{Pruneau06a}  C. Pruneau, {\em et al.}  (STAR Collaboration), J. of Physics G: Nucl. Part. Phys. {\bf34}, S667 (2007).



\bibitem{Berger} E. L. Berger, Nucl. Phys. {\bf B85}, 61 (1975)..

\bibitem{Carruthers} P. Carruthers and I. Sarcevic, Phys. Rev. Lett. {\bf 63}, 1561 (1989).


\bibitem{StarNim} M. Anderson {\it et al.} (STAR Collaboration), Nucl. Instrum. Meth. {\bf A 499} 659 (2003).
\bibitem{StarC308} STAR Collaboration, private communication.

\bibitem{CDF_JET} T. Affolder, {\it et al.} {\em STAR Collaboration}, Phys. Rev. D {\bf 65}, 092002 (2002).

\bibitem{WhitePapers} I.~Arsene {\it et al.}  {\em BRAHMS Collaboration}, Nucl.\ Phys.\ A {\bf 757}, 1 (2005);  K.~Adcox {\it et al.}   
{\em PHENIX Collaboration}, {\it ibid} .~184;   B.~B.~Back {\it et al.} {\em PHOBOS Collaboration},  {\it ibid} .~28;  J.~Adams {\it et al.} {\em STAR Collaboration},
  {\it ibid} . ~102.

\bibitem{StarV4} J. Adams {\it et al.} {\em STAR Collaboration}, Phys. Rev. C {\bf 72}, 014904 (2005).
\bibitem{VoloshinReview08} S. A. Voloshin, A. M. Poskanzer, and R. Snelling, {\em nucl-ex/0809.2949}.
\bibitem{ZYAM} N. N. Ajitanand, {\it et al.} Phys.Rev. C {\bf 72}, 011902(2005).

\bibitem{Horner06}  M. Horner, {\em et al.}  (STAR Collaboration), J. of Physics G: Nucl. Part. Phys. {\bf 34}, S995 (2007).

\end{thebibliography}
\end{document}